\documentclass[twocolumn,superscriptaddress,amsmath,amssymb,aps,prx]{revtex4-1}

\usepackage{graphicx}
\usepackage{dcolumn}
\usepackage{bm}

\usepackage[colorlinks,urlcolor=blue,citecolor=blue,linkcolor=blue]{hyperref}

\newcommand{\vect}[1]{{\mathbf #1}}

\renewcommand{\k}{{\bf k}}
\newcommand{\p}{{\bf p}}
\renewcommand{\P}{{\bf P}}
\newcommand{\q}{{\bf q}}

\newcommand{\0}{{\bf 0}}

\newcommand{\am}{a_-}
\newcommand{\ab}{a_{\rm B}}

\newcommand{\bra}[1]{\left\langle{#1}\right|}
\newcommand{\ket}[1]{\left|{#1}\right>}

\newcommand{\eb}{\varepsilon_B}
\newcommand{\eq}{\epsilon_{\q}}
\newcommand{\ek}{\epsilon_{\k}}

\newcommand{\ekone}{\epsilon_{\k_1}}
\newcommand{\ektwo}{\epsilon_{\k_2}}

\newcommand{\nn}{\nonumber}
\newcommand{\beq}{\begin{equation}}
\newcommand{\eeq}{\end{equation}}
\newcommand{\inv}{^{-1}}

\newcommand{\T}{{\cal T}}

\newcommand{\sch}{Schr{\"o}dinger }

\renewcommand{\ab}{a_{\text{B}}}

\def\lba{\left(}    \def\rba{\right)}
\def\lbc{\left[}    \def\rbc{\right]}

\def \del{\partial}       

\newcommand{\sout}[1]{}

\begin{document}

\title{Universality of an impurity in a Bose-Einstein condensate}

\author{Shuhei M. Yoshida}
\affiliation{Department of Physics, The University of Tokyo, Tokyo 113-0033, Japan}
\affiliation{School of Physics and Astronomy, Monash University, Victoria 3800, Australia}

\author{Shimpei Endo}
\affiliation{School of Physics and Astronomy, Monash University, Victoria 3800, Australia}

\author{Jesper Levinsen}
\affiliation{School of Physics and Astronomy, Monash University, Victoria 3800, Australia}

\author{Meera M.\ Parish}
\affiliation{School of Physics and Astronomy, Monash University, Victoria 3800, Australia}

\date{\today}

\begin{abstract}
Universality is a powerful concept in physics, allowing one to construct physical descriptions of systems that are independent of the precise microscopic details or energy scales.
A prime example is the Fermi gas with unitarity limited interactions, 
whose universal properties are relevant to systems ranging from atomic gases at microkelvin temperatures to the inner crust of neutron stars. 
Here we address the question of whether unitary Bose systems can possess a similar universality.
 We consider the simplest strongly interacting Bose system, where we have an impurity particle 
 (``polaron'') resonantly interacting with a 
 Bose-Einstein condensate (BEC). 
 Focusing on the ground state of the equal-mass system, we use a variational wave function for the polaron that includes up to three Bogoliubov
  excitations of the BEC, thus allowing us to 
  capture both Efimov trimers and associated tetramers. 
  Unlike the Fermi case, we find that the length scale associated with Efimov trimers (i.e., the three-body parameter) can strongly affect the polaron's behaviour, even at boson densities where there are no well-defined Efimov states.
  However, by comparing our results with recent quantum Monte Carlo calculations, we argue that the polaron energy 
  is a \emph{universal} function of the Efimov three-body parameter 
  for sufficiently low boson densities.
We further support this conclusion by showing that the energies of the deepest bound Efimov trimers and tetramers at unitarity are universally related to one another, regardless of the microscopic model.
On the other hand, we find that the quasiparticle residue and effective mass sensitively depend on the coherence length $\xi$ of the BEC, with the residue tending to zero as $\xi$ diverges, in a manner akin to the orthogonality catastrophe.
\end{abstract}


\maketitle

\section{Introduction}
Quantum gases 
are expected to exhibit universal thermodynamic and dynamical properties when the scattering length $a \to \pm \infty$ 
and the interparticle spacing greatly exceeds the range of the interactions. 
Here, the large separation of length scales implies that the unitary gas is independent of an interaction length scale, thus making it insensitive to its microscopic properties. This scenario exists 
for the two-component Fermi gas at unitarity~\cite{FGbook}, 
where highly controlled cold-atom experiments have recently determined  
its universal equation of state \cite{mit,ens,tokyo}, pairing correlations \cite{Partridge_2005,Stewart_2008,Schirotzek_2008,Inada_2008,Veeravalli_2008,Navon2010,Stewart_2010,Kuhnle_2010} 
and transport properties 
\cite{Cao2011}.
Moreover, Bose gases in the unitary regime 
are now also being experimentally probed using ultracold atomic gases \cite{Makotyn2014,Hu2016,jorgensen2016,Fletcher2017,Lopes2017,Klauss2017,Eigen2017}.
However, it remains an open question whether 
the unitary Bose system can also display universal behavior that has relevance to a range of different 
physical systems.

In contrast to the Fermi case, bosonic systems at unitarity always support few-body Efimov bound states~\cite{Efimov1970,Braaten2006,Kraemer2006,naidon2017,greene2017,dIncao2017}, whose energy and size depend on short-distance parameters such as the interaction range $r$. 
Thus, there is an additional interaction length scale that can impact the behavior of the unitary Bose system and potentially make it 
sensitive to microscopic details. 
Indeed, the deepest bound trimers and larger few-body clusters typically 
cannot be universally related to each other without invoking a specific model for the short-range interactions~\cite{naidon2017,greene2017}.

To investigate the universality of a Bose system at unitarity, we consider an impurity immersed in a Bose-Einstein condensate (BEC), where the boson-impurity interactions can be tuned to unitarity, while the rest of the system remains weakly 
interacting. This so-called ``Bose polaron'' has received much theoretical attention~\cite{Tempere2009,Rath2013,Li2014,Shashi2014,Christensen2015,Levinsen2015,Ardila2015,Grusdt2015,GrusdtDemlerReview,Ardila2016,Shchadilova2016,Loft2017,Grusdt2017,Sun2017,Lampo2017} and has very recently been observed in cold-atom experiments~\cite{Hu2016,jorgensen2016}. 
Moreover, it has 
been extended to impurities with more complex internal structure~\cite{Schmidt2015b,Schmidt2016b,Camargo2017}, and it has
potential relevance to polaron problems in solid-state systems --- for instance, the limit of weak impurity-boson interactions can be directly mapped to the Fr\"{o}hlich model \cite{Tempere2009,Rentrop2016}.
The Bose polaron is a promising system in which to search for universal behavior, since the Efimov trimers consisting of the impurity and two bosons are orders of magnitude larger than the short distance scale $r$
(provided the impurity mass is not too small)~\cite{Wang2012b,Blume2014}. 
If we parameterize the size of the deepest Efimov trimer by $|a_-|$, where $a_-$ is the scattering length at which the trimer crosses the three-atom continuum (see Fig.~\ref{fig:fewbody}), then the hierarchy of length scales at unitarity is $r \ll n^{-1/3} \ll |a_-|$ for the typical densities $n$ in experiment~\cite{Hu2016,jorgensen2016}. This suggests that there is a regime where Efimov physics is irrelevant, 
such that the ground-state properties of the polaron only depend on the density, 
like in the case of an impurity resonantly interacting with a Fermi gas~\cite{chevy2006_2,Combescot2008,
Prokofev2008a,
Vlietinck2013,Massignan_Zaccanti_Bruun}. 

In this paper, we address this question using impurities that have a mass equal to that of the bosons ---
a situation which has already been realized in 
the $^{39}$K atomic gas experiments in Aarhus \cite{jorgensen2016}. 
We first investigate the few-body limit, and 
determine the Efimov trimer and associated tetramer states that involve the impurity. 
For vanishing boson-boson interactions,
we show that the ratios of the binding energies for the deepest bound states are universal, unlike the identical-boson case. We then include this Efimov physics in the many-body system by constructing a variational wave function for the ground-state Bose polaron which recovers both three- and four-body equations in the limit of zero density. Strikingly, we find that the polaron properties at unitarity sensitively depend on the Efimov scale $|a_-|$, even in the regime 
where $|\am|$ far exceeds the interparticle spacing.
However, we show 
that the ground-state polaron energy is a \emph{universal} function of $n^{1/3}|a_-|$ that is model independent in the regime $n^{1/3} r \ll 1$. 
We corroborate this finding by comparing our results with recent quantum Monte Carlo (QMC) calculations \cite{Ardila2015,Ardila2016}.

In the non-universal, high-density limit $n^{1/3} r \gg 1$, we
consider the case of a narrow Feshbach resonance and we derive perturbative expressions for the polaron energy and contact at unitarity. This allows us to demonstrate, for the first time, that the Bose polaron at unitarity can be thermodynamically stable even in the limit of vanishing boson-boson interactions. 
On the other hand, we find that the quasiparticle residue and effective mass sensitively depend on the BEC coherence length. For the case of an ideal BEC, the effective mass converges to a finite value as we increase the number of excitations of the condensate, while the residue vanishes in a manner reminiscent of the orthogonality catastrophe for a static impurity in a Fermi gas~\cite{Anderson1967}. 

\section{Model and variational approach}

We consider an impurity atom immersed in a weakly interacting
homogeneous Bose-Einstein condensate at zero temperature. The
interactions in the medium are characterised by the boson-boson
scattering length $\ab$, and we thus require $n_0\ab^3\ll1$, with $n_0$
the condensate density (which essentially equals the total density $n$ in this regime). This allows us to treat both the ground state
and the excitations of the BEC within Bogoliubov
theory~\cite{Fetter}. 
Note that we always implicitly assume $\ab>0$ to ensure the stability of the condensate.

The presence of the impurity in the medium adds two 
length scales associated with the impurity-boson interaction.
The first of these is the scattering length $a$ between the impurity and a boson from the medium. 
We disregard this length scale since we will focus on the strong-coupling unitary regime close to a Feshbach resonance, where
$|a|$ greatly exceeds all other length scales, i.e., we generally
consider $1/a=0$.  Here, we assume short-range contact interactions, which is well realized in current cold-atom experiments.

The second additional length scale is the so-called three-body
parameter, which characterizes the size of the smallest (ground-state) Efimov trimer consisting of two bosons and the impurity. 
To demonstrate the model independence of our results, we fix this Efimov length scale in two different ways:
\begin{itemize}
\item $r_0$-\textit{model} --- we introduce  an effective range $r_0$, which directly relates the three-body parameter to the two-body interaction~\cite{Petrov2004tbp}
\item $\Lambda$-\textit{model} --- 
we take $r_0 \to 0$ and 
instead apply an explicit ultraviolet 
cutoff $\Lambda$ to the momenta involved in Efimov physics, which is equivalent to including a three-body repulsion~\cite{Bedaque1999}
\end{itemize}
\noindent
As we explicitly show in Section~\ref{sec:few-body}, 
for an impurity of the same mass as the
medium atoms --- the scenario considered here --- the precise manner of regularizing Efimov physics is unimportant owing to the large separation of scales between the short-range physics and Efimov physics.

Given the above considerations, we model the system 
using a two-channel description of the Feshbach resonance~\cite{Timmermans1999fri}, which corresponds to the
Hamiltonian (setting $\hbar$ and the volume to 1):
\begin{align}\label{eq:Ham2}
  \hat{H} = & \sum_\k \left[ E_\k \beta^\dag_\k \beta_\k +
              \epsilon_\k
              c^\dag_\k c_\k + 
              \left( \epsilon^{\rm d}_\k + \nu_0 \right)
              d^\dag_\k d_\k \right ] \nn  \\ & \hspace{-10mm}
+g \sqrt{n_0} \sum_\k \left( d^\dag_\k c_\k+ h.c. \right)
+ g \sum_{\k,\q } \left(d^\dag_\q 
                                                           c_{\q -\k}
b_\k + h.c.   \right).
\end{align}
Here, $b^\dag_\k$ and $c^\dag_\k$ are the creation
operators of a boson and the impurity, respectively, with momentum
$\k$ and single particle energy $\ek=\k^2/2m$, where the mass $m$ of a
boson and of the impurity are taken to be equal. A boson and the impurity
interact by forming a closed-channel dimer created by $d^\dag_\k$,
with $\epsilon^{\rm d}_\k=\ek/2$.  The inter-channel coupling $g$ and
the bare detuning $\nu_0$ are chosen such that they reproduce the
two-body scattering amplitude in vacuum
at relative momentum $k$:
\begin{align}
f(k)=-\frac1{a^{-1}-\frac12r_0k^2+ik}.
\end{align}
Carrying out the renormalization procedure with high momentum cutoff $k_0$, one obtains~\cite{Bruun2004,Levinsen2011}
\begin{align} \label{eq:observables}
a=\frac{mg^2}{4\pi}\frac1{\frac{g^2m k_0}{2\pi^2}-\nu_0}, \hspace{10mm}
r_0=-\frac{8\pi}{m^2g^2},
\end{align}
which relates the physical low-energy observables --- the scattering length
$a$ and effective range $r_0\leq0$ --- to the bare parameters of the model. In all calculations, we take the limit $k_0, \nu_0 \to \infty$ for a given set of values for the observables.
We recover the single-channel model (where $r_0 = 0$) by taking $g \to \infty$ while holding $a$ fixed.

In writing the Hamiltonian \eqref{eq:Ham2} we already applied the
Bogoliubov approximation for the weakly-interacting condensate: The
operator $\beta^\dag_\k$ creates a Bogoliubov excitation,
and is related to the bare boson operator $b^\dag_\k$ by the
transformation $b^\dag_\k=u_\k \beta^\dag_\k - v_\k
\beta_{-\k}$.
Here, $u_\k=\sqrt{[(\ek+\mu)/E_\k+1]/2}$,
$v_\k=\sqrt{[(\ek+\mu)/E_\k-1]/2}$, the Bogoliubov dispersion is
$E_\k=\sqrt{\ek(\ek+2\mu)}$, and the boson chemical potential is
$\mu = 4\pi n_0 \ab/m \equiv 1/(2 m \xi^2)$, where $\xi$ is the coherence length of the condensate.
Note that Eq.~\eqref{eq:Ham2} is defined with respect to the energy of the BEC in the absence of the impurity. 

To explore the ground-state properties of the unitary Bose polaron, we apply the variational principle, using a wave function
consisting of terms $\ket{\psi_N}$ with an increasing number  $N$ of Bogoliubov modes excited from
the condensate by the impurity:
\begin{align}\label{eq:varWF}
\ket{\Psi}=\ket{\psi_0}+\ket{\psi_1}+\ket{\psi_2}+\ket{\psi_3}.
\end{align}
Explicitly, we have
\begin{align}
\ket{\psi_0} = & 
    \alpha_0 c^\dag_{0} \ket{\Phi}\nn \\
\ket{\psi_1} = & 
\left(\sum_\k \alpha_\k c^\dag_{- \k} \beta^\dag_\k
 + \gamma_0 d^\dag_0  \right) \ket{\Phi}\nn \\
\ket{\psi_2} = & 
\left(\frac{1}{2} \sum_{\k_1 \k_2} \alpha_{\k_1 \k_2} 
        c^\dag_{- \k_1 - \k_2} \beta^\dag_{\k_1} \beta^\dag_{\k_2} 
    + \sum_\k \gamma_\k d^\dag_{- \k} \beta^\dag_\k 
\right) \ket{\Phi}\nn \\
\ket{\psi_3} = &   \Bigg(\frac{1}{6} \sum_{\k_1 \k_2 \k_3} \alpha_{\k_1 \k_2 \k_3}
        c^\dag_{-\k_1-\k_2-\k_3} \beta^\dag_{\k_1} \beta^\dag_{\k_2}
                 \beta^\dag_{\k_3}\nn \\ &
    \hspace{16mm}+ \frac{1}{2}\sum_{\k_1 \k_2} \gamma_{\k_1 \k_2} d^\dag_{-\k_1-\k_2}
        \beta^\dag_{\k_1} \beta^\dag_{\k_2}
                  \Bigg) \ket{\Phi},\nn \\ &
\label{eq:psi}
\end{align}
with $\ket{\Phi}$ the ground state wave function of the weakly
interacting BEC. Including up to three excitations allows us to investigate the effect of \textit{both} Efimov trimers and tetramers on the unitary Bose polaron.
This goes beyond
previous works on the Bose polaron: The effect of dressing the impurity by a single excitation
was investigated in Ref.~\cite{Rath2013} using a $T$-matrix approach
and in Ref.~\cite{Li2014} with a variational approach. Furthermore, two of us previously used two-excitation dressing to investigate the relationship between
polaronic and Efimov physics~\cite{Levinsen2015}, as well as to successfully compare 
with the experimentally obtained
polaron spectral function~\cite{jorgensen2016} (see also Ref.~\cite{Parish2016}).

To investigate the properties of the ground-state Bose polaron, we determine the variational parameters by the stationary condition
$\partial_{\alpha^\ast,\gamma^\ast}\bra{\Psi}(\hat{H}-E)\ket{\Psi}=0$,
where the energy $E$ can be viewed as the Lagrange multiplier ensuring normalization.  This
yields a set of coupled equations, from which we can eliminate the
$\alpha$ coefficients and obtain coupled integral equations for the
$\gamma$'s only. The resulting equations are given in
Appendix~\ref{app:inteqs}. In the $r_0$-model, we keep the effective range $r_0$ in the two-body interaction, while in the $\Lambda$-model, we set $r_0=0$ and apply the cutoff $\Lambda$ to the momenta in the $\gamma$ coefficients.

\section{Universal few-body Efimov states}
\label{sec:few-body}

\begin{figure*}[t]
\centering
\includegraphics[width=1.3\columnwidth]{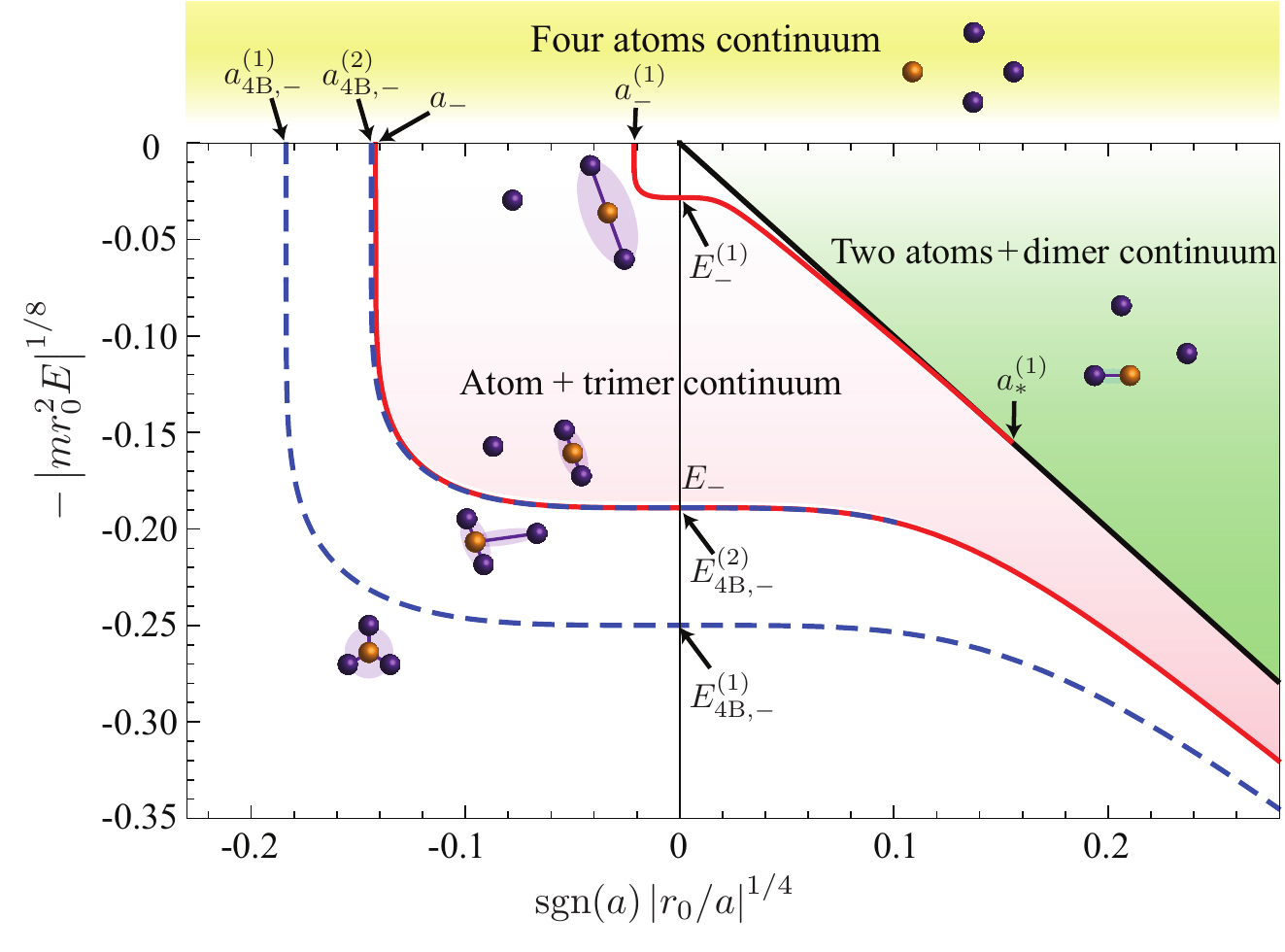}
\caption{Few-body energy spectrum calculated from 
equations (\ref{eq:2body}-\ref{eq:4body}). The ground (first excited) trimer
is shown as a red solid curve, which first 
appears at a negative scattering length $a_{-}$ $(a_{-}^{(1)})$ and then dissociates into the atom-dimer continuum (delimited by the 
black solid line) at a positive scattering length $a_{*}$ $(a_{*}^{(1)})$, where $a_*$ is outside the plotted region. We only show the ground and first excited trimers, but there exists an infinite series of higher excited trimers 
with an accumulation point at unitarity, $1/a=0$. We also find two tetramer states (blue dashed lines) tied to the ground Efimov trimer. The excited tetramer is very weakly bound, but it persists at unitarity and disappears into the atom-trimer continuum at the positive scattering length $a\simeq 9\times10^{3}|r_0|$. }
\label{fig:fewbody}
\end{figure*}

Before proceeding to the many-body polaron physics, it is important to
first discuss the few-body limit since --- as we shall explicitly
demonstrate in Section~\ref{sec:polaron} 
--- this plays a crucial
role in understanding the polaron properties.
To determine the few-body spectrum, we use wave functions of the form 
\eqref{eq:psi} in the limit $n\to0$, such that $\ket{\Phi}$ now
corresponds to the Fock vacuum. Here, the state
$\ket{\psi_N}$ with $N$ excitations is identified as a few-body state containing $N$ bosons and the impurity. Consequently, the
varational approach is exact and the sectors of different particle
numbers completely decouple. Ignoring the trivial equation arising
from the single-particle problem, the resulting few-body equations
take the form
\begin{align}
\T\inv_0(E,\0) & =  0,
\label{eq:2body}  \\
\T\inv_0(E-\epsilon_\k,\k) \, \gamma_\k & =  \sum_{\q}
    \frac{\gamma_\q}{E-\epsilon_{\k\q}},
\label{eq:3body}  \\
 \T\inv_0(E-\epsilon_{\k_1}-\epsilon_{\k_2},\k_1+\k_2) \gamma_{\k_1 \k_2} & = 
 \sum_{\q} 
    \frac{\gamma_{\k_1 \q}+\gamma_{\q \k_2 }}{E-\epsilon_{\k_1 \k_2 \q}},
\label{eq:4body}
\end{align}
which may be obtained as the zero-density limit of the polaron equations in Appendix~\ref{app:inteqs}. 
Here we have defined $\epsilon_{\k_1\k_2}\equiv
\epsilon_{\k_1}+\epsilon_{\k_2}+\epsilon_{\k_1+\k_2}$
and $\epsilon_{\k_1\k_2\q}\equiv
\epsilon_{\k_1}+\epsilon_{\k_2}+\epsilon_{\q}+\epsilon_{\k_1+\k_2+\q}$, while
$\T_0$ is the vacuum two-body $T$-matrix, 
\begin{align}
    \T_0\inv(E,\k)
    &= \frac{m}{4\pi a} - \frac{m^2 r_0}{8\pi} (E-\epsilon^d_{\k})
        - \frac{m^{\frac32}}{4\pi}\sqrt{\epsilon^d_\k -E-i0}.
\label{eq:tmatrix}
\end{align}
Equation~\eqref{eq:2body} gives the condition for a two-body bound state, while the solutions of Eqs.~\eqref{eq:3body} and \eqref{eq:4body} yield, respectively, the trimer spectrum 
and associated tetramers~\cite{Pricoupenko2011}.
Note that boson-boson interactions are neglected in this limit, since they are only included via a density-dependent mean-field shift in the boson chemical potential. However, this approximation is reasonable when
the Efimov states are insensitive to $\ab$, i.e., when the microscopic length scale determining the three-body parameter greatly exceeds $\ab$.

In Figure~\ref{fig:fewbody}, we show the spectrum of bound states obtained by solving
Eqs.~(\ref{eq:2body}--\ref{eq:4body}) within the $r_0$-model. 
For positive scattering length,
a single two-body bound dimer exists, with energy
\begin{align}
\eb=-\frac{\left(\sqrt{1-2r_0/a}-1\right)^2}{mr_0^2},
\label{eq:dimer}
\end{align}
as found by solving
Eq.~\eqref{eq:2body}. The three-body problem described by
Eq.~\eqref{eq:3body} is distinctly different: For strong interactions,
the impurity and two bosons can form an infinite series of three-body
bound states, even in the absence of the
two-body bound state. Remarkably, these so-called Efimov
trimers~\cite{Efimov1970,Efimov1973} satisfy a discrete scaling symmetry whereby the energy $E$ of one trimer can be mapped onto the next via
the discrete transformation $E\to \lambda_0^{-2l}E$ and
$a\to\lambda_0^la$, with $l$ integer and $\lambda_0$ a scaling
parameter that depends on the particular three-body problem (for reviews, see Refs.~\cite{Braaten2006,naidon2017,greene2017,dIncao2017}).
In our case
of two non-interacting identical bosons with the same mass as the
third particle, the scaling factor is found from a transcendental
equation to be
$\lambda_0=e^{\pi/s_0}\simeq1986.1$~\cite{Efimov1973}. 
Due to this large scaling
factor, we only show the two lowest trimer states in the figure
(indeed, for realistic experimental parameters, the size of the first excited Efimov trimer is of order 1cm!).

Below the ground-state Efimov trimer, we predict the existence of two tetramer states, even at unitarity (see Fig.~\ref{fig:fewbody}). 
This mirrors previous results for four
identical bosons, where it was found, both theoretically~\cite{hammer2007,Stecher2008,PhysRevLett.107.135304,deltuva2011} and experimentally~\cite{Ferlaino2009}, 
that there are two lower-lying tetramers associated with the
ground-state trimer. The present scenario more closely resembles studies of 
a light atom resonantly interacting with three identical heavy
atoms~\cite{Blume2014,PhysRevLett.108.073201,Schmickler2017}. Our findings are generally consistent with the results of
these previous works, although no excited tetramer 
was reported for small mass imbalance~%
\footnote{We have not investigated tetramer states tied to the excited Efimov trimer, but we expect that they will exist as resonant states embedded in the atom-trimer continuum, as is the case for identical bosons~\cite{deltuva2011,Stecher2011PRL}.}.
Note that, within our theory, we do not require an additional length scale to fix the positions of the tetramers relative to the Efimov trimers~\cite{PhysRevLett.107.135304}.

\begin{table*}[t]
\renewcommand{\arraystretch}{1.8}
\begin{tabular}{|c||c|c|c|c|c|c|c|c|}
\hline
& $|a_-|$ & $ma^2_-|E_-|$ & $a_-^{(1)}/a_-$ & $\sqrt{E_-/E_-^{(1)}}$ &
$a_-/a_{\mathrm{4B},-}^{(1)}$ & $E_{\mathrm{4B},-}^{(1)}/E_-$ &
$a_-/a_{\mathrm{4B},-}^{(2)}$ & $E_{\mathrm{4B},-}^{(2)}/E_-$ \\
\hline
$r_0$-model & $2467|r_0|$ & 9.913 & 1991 & 1986.126 & 2.810 & 9.35 & 1.060 & 1.0030(3)\\
$\Lambda$-model & $1354\Lambda^{-1}$ & 9.950 & 1987 & 1986.127 & 2.814 & 9.43 & 1.061 & 1.0036(1) \\
\hline
\end{tabular}
\caption{Comparison of few-body data for the $r_0$- and the $\Lambda$-model. While the overall length scale set by $a_-$ depends on the details of the models, the dimensionless ratios characterizing the few-body spectra agree to an accuracy of $1\%$ or less. These quantities include the scattering lengths at which the few-body bound states cross into the continuum, as well as their energies at unitarity, and are illustrated in Fig.~\ref{fig:fewbody}.\label{tab:efimov} }
\end{table*}

A crucial point is that the few-body physics described here is essentially \textit{universal}, in the sense that the entire spectrum is independent
of the details of the short-range physics. This universality
originates from the large separation of scales between the short-distance physics, characterized here by the effective range $r_0$, and the typical size of the ground-state Efimov trimer. The latter may be conveniently characterized by the ``critical'' scattering length $a_-$ at which the trimer crosses into the three-atom continuum. In our model, we have $a_-/r_0\simeq2467$~\footnote{This number had a typo in Ref.~\cite{Levinsen2015}  which did not affect any of the other results of that paper.} (see Table \ref{tab:efimov} for a summary of the length scales and energy scales associated with Efimov physics). This large ratio ensures that the energy of the ground-state Efimov trimer at unitarity is far smaller than that required to probe the short-range physics. Indeed, we see that the ratio of critical scattering lengths for
the excited and 
ground-state trimers, 
i.e., $a_-^{(1)}/a_-\simeq1991$, is very close to the predicted scaling factor of $1986.1$. Likewise, at unitarity, the ratio of the ground-state trimer energy to that of the excited state is 
$E_-/E_-^{(1)}\simeq(1986.1)^2$, i.e., it matches the universal prediction to 5 significant digits. 
This is in remarkable contrast to the situation for three identical bosons, where deviations from the universal scaling factor of 22.7 are usually on the order of 10-20\% due to finite-range corrections~\cite{PhysRevLett.112.190401,Braaten2006,naidon2017}.

To test the model independence of the few-body physics, we also consider the $\Lambda$-model, where $r_0=0$ and we apply an explicit cutoff $\Lambda$ to the momentum sums in the three- and four-body equations (\ref{eq:3body}--\ref{eq:4body}).
Here we find that after fixing the three-body parameter, the spectrum is essentially indistinguishable from that in Fig.~\ref{fig:fewbody}, in the sense that the differences are within the thickness of the lines~\footnote{The exception to this expected universality is the crossing of the ground-state trimer with the atom-dimer continuum (not shown in Fig.~\ref{fig:fewbody}). This crossing is in general known to be the least universal aspect of Efimov trimers, as it corresponds to the largest energy scale. We indeed find that the ground state trimer crossing is at $a_*/|a_-|\simeq4.86\times 10^{-4}$ in the $r_0$-model and at $a_*/|a_-|\simeq5.58\times 10^{-4}$ in the $\Lambda$-model, neither of which quite fits the universal prediction: $a_*/|a_-|\simeq0.72418e^{-\pi/s_0}=3.6\times10^{-4}$ \cite{helfrich2010}. On the other hand, we find that the first excited trimer, with $a_*^{(1)}/|a_-|\simeq0.71$, matches reasonably well.
}. The agreement between the two models is also evident in the summary of length scales and energy scales in Table~\ref{tab:efimov}.
Thus, we conclude that our description of the few-body spectrum can be accurately characterized by universal Efimov theory.

Further evidence of the universality of our few-body results is found when comparing the ratio of our ground-state tetramer and trimer energies with recent 
 QMC results~\cite{Ardila2015}. There, this ratio was reported to be $E_{\mathrm{4B},-}^{(1)}/E_-\simeq 9.7$, while we find   $E_{\mathrm{4B},-}^{(1)}/E_-\simeq9.35$. 
Importantly, the QMC calculation used a completely different model involving a hard-sphere boson-boson interaction, characterized by the scattering length $\ab$, and
attractive square-well interactions between the impurity and the bosons. The range of the square well was taken to be much smaller than $\ab$, and consequently the energies of the 
ground-state Efimov trimer and associated tetramers are set by $\ab$. The universality of the spectrum thus means that we can relate the three-body parameter in the QMC calculation to the boson-boson scattering length 
\footnote{Under the assumption of universality where the quantity $m\am^2E_{\mathrm{4B},-}^{(1)}$ is the same in all models, 
the ratio is given by $a_-^{\mathrm{{\scriptscriptstyle (QMC)}}} / \ab^{\rm{\scriptscriptstyle (QMC)}}
\simeq\sqrt{r_0^2 E_{\mathrm{4B},-}^{(1,{\scriptscriptstyle 2ch})} / \ab^2 E_{\mathrm{4B},-}^{(1,{\scriptscriptstyle QMC})}} 
a_-^{({\scriptscriptstyle 2ch})} 
/ r_0^{({\scriptscriptstyle 2ch})}$. All parameters on the right-hand-side are known from the results of the QMC and the $r_0$-model.}:
\begin{align} \label{eq:qmc}
a_-^{\mathrm{{\scriptscriptstyle (QMC)}}}\simeq- 2.09(36) \times 10^4 \ab^{\rm{\scriptscriptstyle (QMC)}}.
\end{align}
The agreement between the results from the different models also reinforces the point that we do not need a four-body parameter to fix the tetramer relative to the trimer. 
This is consistent with theoretical results for identical bosons \cite{hammer2007,Stecher2008,deltuva2011} and for heteronuclear systems~\cite{Blume2014,PhysRevLett.108.073201,Schmickler2017}. 

We end this section by contrasting 
the few-body universality found here with the so-called van der Waals universality of Efimov physics~\cite{Berninger2011,Wang2012a,naidon2014a,naidon2014b,wang2014}. The latter refers to the phenomenon where the Efimov states of various atomic species 
are universally related to
their van der Waals length, e.g., $a_{-}\simeq-(9.1 \pm 1.5) r_{\mathrm{vdW}}$. This behavior, however, is rather different from that predicted by  universal zero-range Efimov theory, in particular for the ground-state trimer.
Indeed, the ratio $a_{-}^{(1)}/a_{-}\simeq20.7$ found in van der Waals universal theory~\cite{wang2014} significantly deviates from the universal ratio 22.7, in contrast to the scenario considered here. 

\begin{figure*}[t]
\centering
\includegraphics[width=1.3\columnwidth]{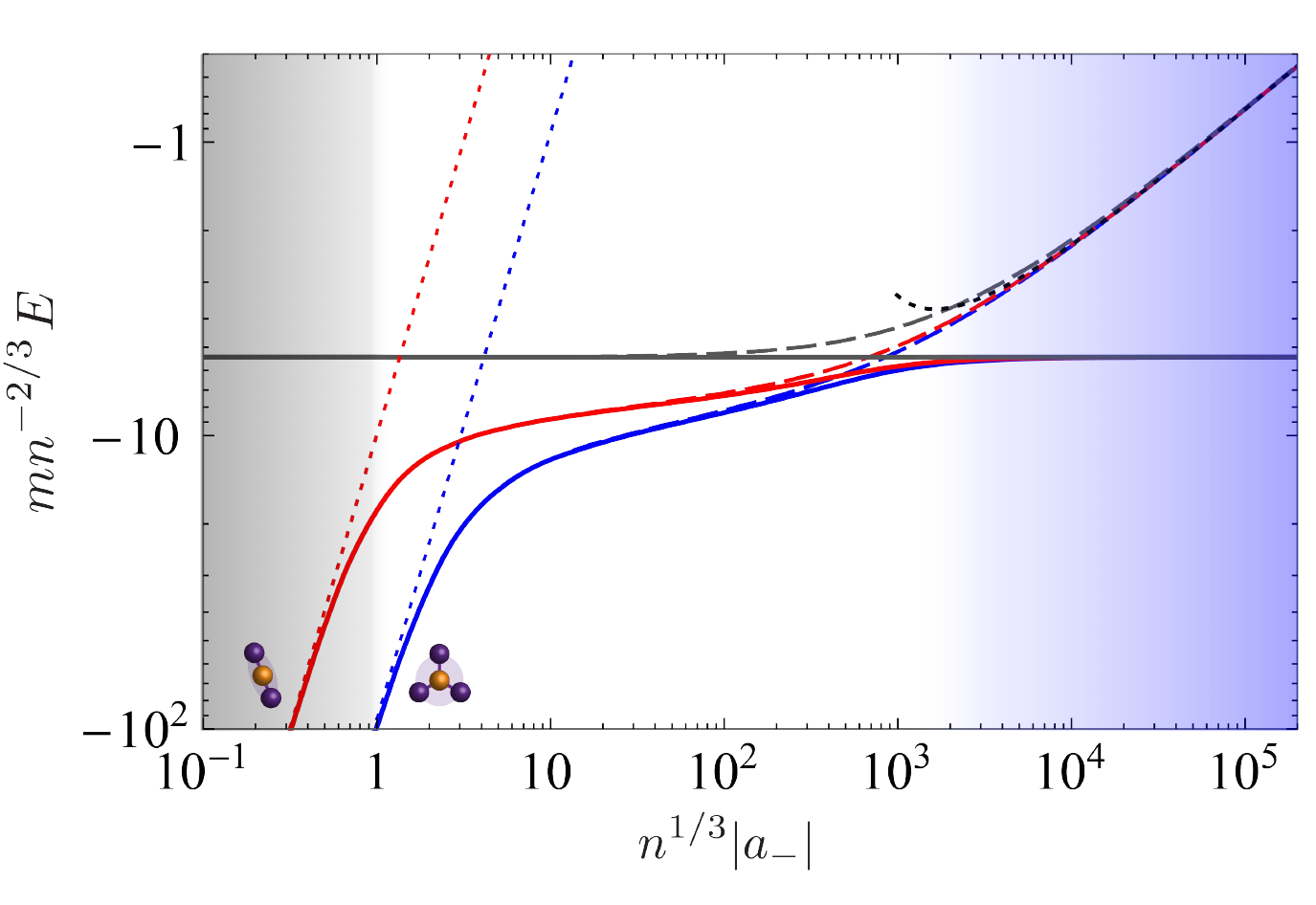}
\caption{Energy equation of state for 
the Bose polaron at unitarity $1/a=0$, obtained using the variational wave function with one (gray), two (red) and three (blue) Bogoliubov excitations for vanishing boson-boson interactions $\ab\to 0$.
We show the results of the $r_0$-model (dashed) and the $\Lambda$-model (solid).
In the low-density limit, the dotted straight lines correspond to the Efimov trimer and associated tetramer energies, with $E_- = - 9.91/m|a_-|^2$ and $E_{\mathrm{4B},-}^{(1)} = - 92.7/m|a_-|^2$, respectively.
The black dotted line in the high-density regime shows the result of a perturbative expansion, Eq.~\eqref{eq:epol}. 
The gray shaded region depicts the low-density few-body dominated regime, while the blue shaded region corresponds to the high-density regime $|r_0|, \Lambda^{-1} > n^{-1/3}$,  where Efimov physics is suppressed and the system is sensitive to the microscopic model.
}
\label{fig:universal}
\end{figure*}

\section{Polaron energy at unitarity}
\label{sec:polaron}
We now turn to the many-body system and investigate the zero-temperature equation of state for the polaron at unitarity.
Since we wish to properly describe Efimov physics within our models, we will focus on the regime where the boson-boson scattering length $\ab$ is always much smaller than all other length scales. 
To test the accuracy of our variational approach outlined in  Eqs.~\eqref{eq:varWF} and \eqref{eq:psi}, we will consider 
wave functions with up to one, two or three Bogoliubov excitations, 
corresponding to $\ket{\psi_0} + \ket{\psi_1}$, $\ket{\psi_0}+ \ket{\psi_1} + \ket{\psi_2}$, and $\ket{\Psi}$, respectively.
For each model, we will convert the short-distance scale (e.g., $r_0$ or $\Lambda^{-1}$) into the dimensionless three-body parameter $n^{1/3}|a_-|$ in order to expose the role of universal Efimov physics. 
The numerical results are displayed in Fig.~\ref{fig:universal} and, in the following, we analyze the different regimes defined by particle density.
  
\subsection{Low-density limit}
\label{sec:lowdensity}  
  
In the limit of vanishing density $n^{1/3}|a_-| \to 0$, the ground-state polaron energy should reduce to that of the deepest bound few-body cluster. If the polaron variational wave function only includes one Bogoliubov excitation, 
  the largest few-body state it can describe is the dimer, which has zero binding energy at unitarity. Thus, as shown in Fig.~\ref{fig:universal}, we obtain an energy that simply tends to zero with decreasing density, i.e., $E=-(4\pi n)^{2/3}/m\simeq-5.4 n^{2/3}/m$ \cite{Rath2013,Li2014}. Note that this is independent of the microscopic model we use since the interparticle spacing greatly exceeds any length scale of the impurity-boson interactions. 
  
  For polaron wave functions with up to two (three) Bogoliubov excitations, the low-density limit recovers the deepest bound universal trimer (tetramer) state discussed in Sec.~\ref{sec:few-body}. Here, the density drops out of the problem and we have polaron energy $E = - \frac{\eta}{m |a_-|^2}$, where $\eta$ is a universal, model-independent constant that depends on the Efimov cluster size (see Fig.~\ref{fig:universal}).

 We also expect there to be larger bound clusters beyond the tetramer, similarly to the case of identical bosons where larger clusters associated with the Efimov effect have been predicted~\cite{Stecher2010JPhys,gattobigio2014,yan2015}
and, indeed, experimentally observed~\cite{zenesini2013}. For the impurity case, recent QMC calculations already demonstrate the existence of a pentamer with $\eta \simeq 300$~\cite{Ardila2015}. However, since there is 
a high-energy cutoff in the problem (e.g., set by $\Lambda$ or $r_0$), the system cannot support arbitrarily deeply bound Efimov clusters, and the deepest bound state should be universal as long as $\eta \ll 10^6$. Indeed, as suggested by the QMC calculations~\cite{Ardila2015}, it is possible that there is no bound cluster larger than the pentamer for the equal-mass impurity case. 
Firstly, the pairwise attraction between the impurity and the bosons scales with $N$ rather than $N(N-1)$, unlike the case for $N$ identical bosons~\cite{greene2017}, so larger bound states will be less favored. Secondly, an effective short-range repulsion between bosons could mean that the pentamer has a closed-shell structure, where any additional bosons must occupy higher energy orbitals, similar to what has been argued for mass-imbalanced fermions~\cite{Bazak2017,Bazak2017PRA}. Such an effective repulsion naturally emerges in both the $r_0$- and the $\Lambda$-models: In the former, this arises from the fact that only one boson at a time can occupy the closed-channel dimer,
while in the latter it is due to the three-body cut-off being equivalent to a three-body repulsion~\cite{Bedaque1999}.

From the above considerations, it is reasonable to conclude that the low-density limit is universal. Moreover, since there is a maximum size to the deepest bound Efimov cluster, we expect the polaron energy to have a well-defined thermodynamic limit, where $mE/n^{2/3}$ is finite for non-zero density. 

\subsection{High-density limit}

  To further support the claim that the polaron energy is finite at unitarity, we consider the opposite limit of high density, $n \to \infty$. 
  Here, the interparticle spacing becomes smaller than the range of the interactions and thus the system will depend on the microscopic details of the model. However, in the case of the $r_0$-model, it allows us to perform a controlled perturbative expansion in the inter-channel coupling $g$ (or, equivalently, $1/n^{1/3}|r_0|$) at unitarity~\cite{Gurarie2007}. Physically, this corresponds to the limit of a narrow Feshbach resonance, a scenario which has already been successfully investigated for an impurity in a Fermi sea, both theoretically~\cite{Massignan2012,Trefzger2012,QiPolaron2012} and experimentally~\cite{Kohstall2012,Cetina2016}.
  
\begin{figure}[ht]
\centering
\includegraphics[width=.9\columnwidth]{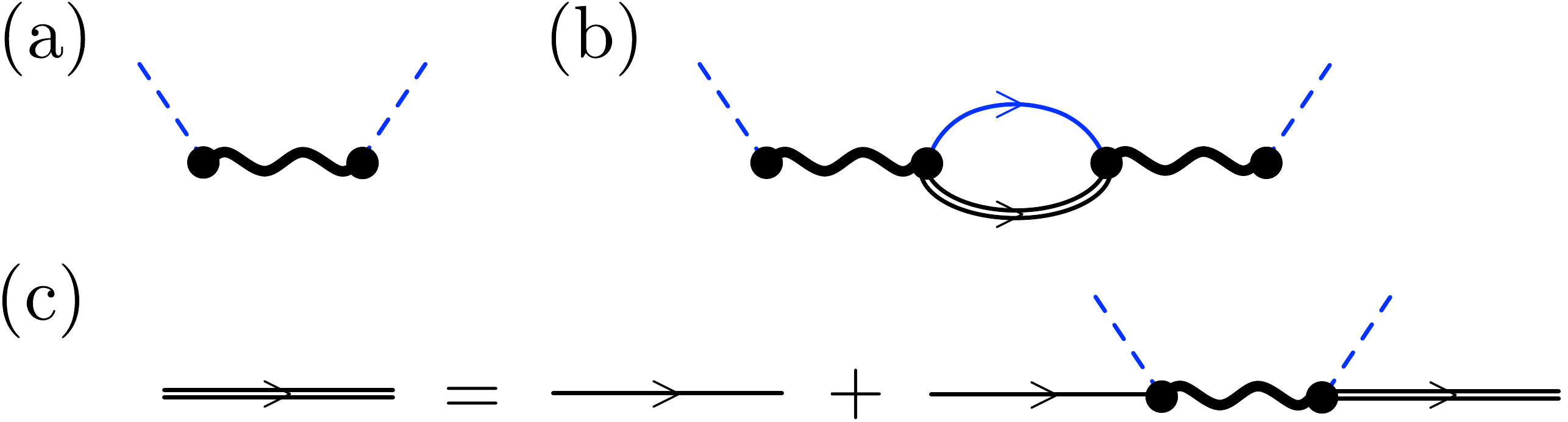}
\caption{Lowest order diagrams for the unitary ground-state polaron 
in the high-density limit $n^{1/3} |r_0| \gg 1$, where (a) and (b) give, respectively, the first and second order terms 
for the polaron energy in a weakly interacting BEC.
Here, the dashed lines correspond to particles emitted or absorbed from the condensate, while the blue solid line is the 
boson propagator, 
the wavy lines represent closed-channel molecule propagators, and the filled circles correspond to the inter-channel coupling $g$. The black double line, defined in (c), represents the dressed impurity to lowest order, where the black single line is the bare impurity propagator.}
\label{fig:diagrams}
\end{figure}

  Our starting point is the self-energy of the unitary Bose polaron, which, in the limit $g \to 0$ and $\ab \to 0$, consists of the lowest order diagrams shown in Fig.~\ref{fig:diagrams}. 
  Such diagrams are included in the self-consistent $T$-matrix approach to the Bose polaron~\cite{Rath2013}.
  For an impurity with momentum $\p$ and frequency $\omega$, 
  the self-energy in this limit is explicitly
  \begin{align} \nn
  \Sigma(\p,\omega) & \simeq  \frac{n_0g^2}{\omega -\epsilon_\p^d} + \frac{n_0 g^4}{(\omega -\epsilon_\p^d)^2}  \times \\  \label{eq:sigma}
  &  \sum_\k \left[\frac{1}{\omega-\epsilon_{\k+\p} -\ek -\frac{n_0g^2}{\omega-\epsilon_{\k+\p}^d-\ek}} + \frac{1}{2\ek} \right],
  \end{align}
  where the first and second terms correspond to the diagrams in Fig.~\ref{fig:diagrams}(a) and (b), respectively.
  The ground-state polaron energy is then determined by taking the zero-momentum pole of the impurity propagator, which gives  $E =  \Sigma(\0,E)$.  Note that we must include self-energy insertions in the impurity propagator even at lowest order (see Fig.~\ref{fig:diagrams}),  since the
  we cannot simply take $\omega = 0$ at the pole when $\ab \to 0$. This is in contrast to previous perturbative treatments of the weakly interacting Bose polaron which focussed on $\ab >0$~\cite{Christensen2015}.
  
In the regime of weak boson-boson interactions, where $n \ab^3   \ll \ab/|r_0|\ll1$,
  we obtain the ground-state polaron energy
  \begin{align}
  E \simeq - \frac{1}{m} \sqrt{\frac{8\pi n}{|r_0|}} + \frac{1}{m}\sqrt{\frac{3}{7}} \left(\frac{8\pi n}{|r_0|^5}\right)^{1/4}.
  \label{eq:epol}
  \end{align}
 This demonstrates that the energy is well-defined and bounded from below in this limit, even when the boson-boson interactions are vanishingly small. 
 Such behavior arises from the fact that only one boson at a time can scatter into the closed channel, thus producing   
 an effective boson-boson repulsion that 
 restricts the density of bosons that can cluster around the impurity.
 As shown in Fig.~\ref{fig:universal}, the perturbative expression correctly reproduces the energy for the $r_0$-model in the high-density limit.
Note that the one-excitation wave function only captures the leading order term in Fig.~\ref{fig:diagrams}(a), while the wave functions with two or three excitations correctly describe the next order correction in $1/n^{1/3}|r_0|$.

In the case of the $\Lambda$-model, the high-density limit corresponds to $n^{1/3}/\Lambda \to \infty$, which is equivalent to taking the cutoff $\Lambda$ to zero. Therefore, within our variational approach, only two-body correlations survive and we obtain the equations for the wave function with one excitation (see Appendix~\ref{app:inteqs}). For vanishing boson-boson interactions, this yields $E = -(4\pi n)^{2/3}/m$, which is the polaron energy of the one-excitation wave function across all densities, as shown in Fig.~\ref{fig:universal}.

\subsection{Many-body universal regime}
\label{sec:manybodyuniversal}
In the intermediate regime $r\ll n^{-1/3}\ll |\am|$ (where $r$ can represent $|r_0|$ or $\Lambda^{-1}$), the interparticle distance is well separated from all length scales associated with the interactions. If the medium were fermionic rather than bosonic, then the polaron energy at unitarity would be a universal value that only depends on the medium density, i.e., $E_{\rm pol}\simeq -4.6 n^{2/3}/m$~\cite{chevy2006_2}. Remarkably, for the Bose polaron, we see in Fig.~\ref{fig:universal} that the energy strongly depends on $n^{1/3}|\am|$ at intermediate densities.
This suggests that there exist resonant three-body interactions, such that there are strong three-body correlations in the system even when the trimer binding energy is comparatively small.

\begin{figure}[ht]
\centering
\includegraphics[width=\columnwidth]{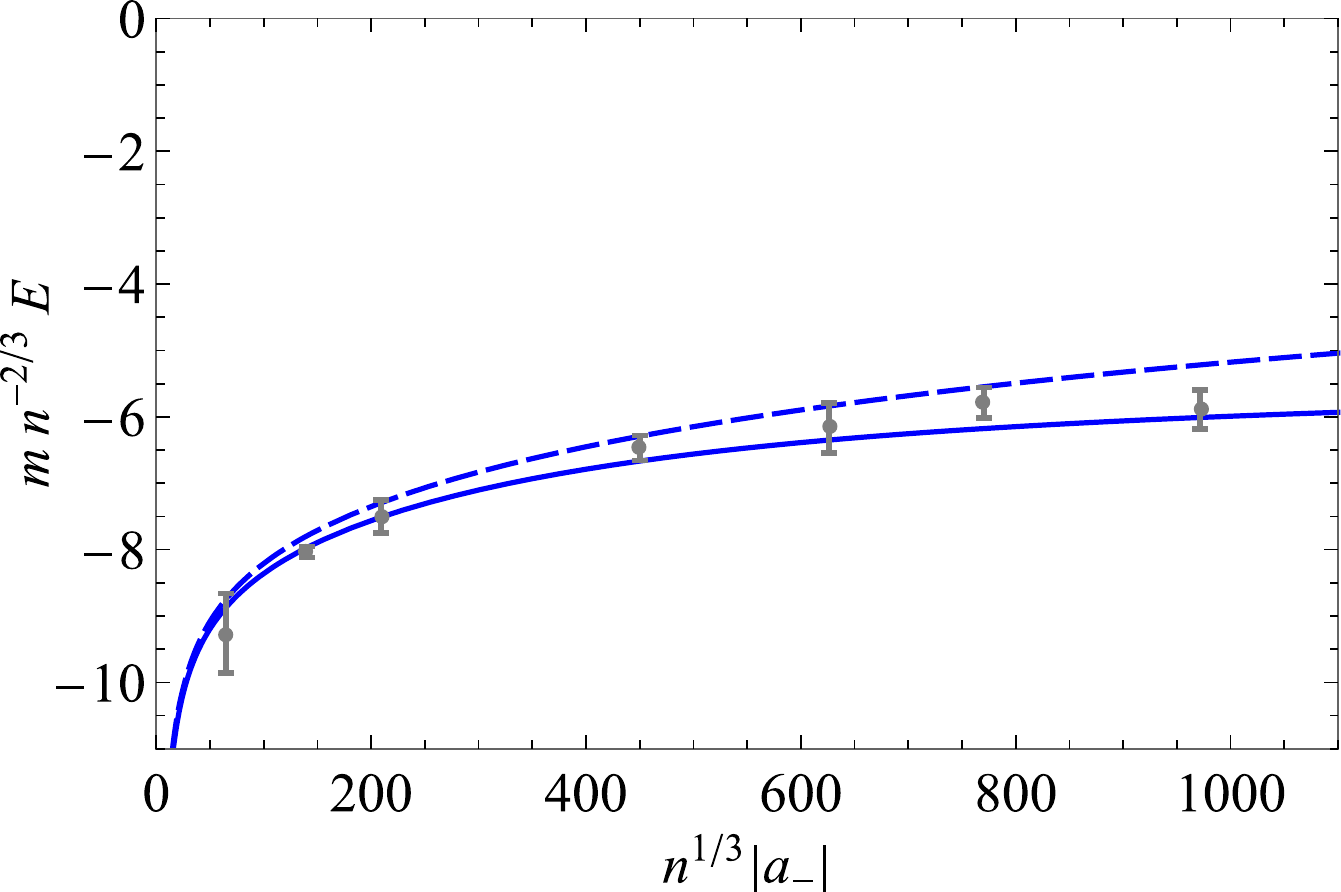}
\caption{Comparison of different models 
for 
the unitary Bose polaron in the regime $r\ll n^{-1/3}\ll |\am|$, where $r$ is the range of the interactions.
We show the results for the polaron energy from the $\Lambda$-model (solid) and the $r_0$-model (dashed), which include up to three excitations of the condensate and which take $\ab \to 0$. 
The QMC results of Ref.~\cite{Ardila2016} are shown as the gray dots.
 The recent Aarhus experiment~\cite{jorgensen2016} is estimated to have $n^{1/3}|a_-| \simeq 100$.
}
\label{fig:comparison}
\end{figure}

Even though the energy of the Bose polaron cannot be assigned a universal value at unitarity, we argue that it is, in fact, a \textit{universal function} of $n^{1/3}|a_-|$ away from the high-density regime. We have previously argued that the low-density limit of the polaron energy universally depends on $n^{1/3}|a_-|$ due to the universal behavior of the few-body states demonstrated in Sec.~\ref{sec:few-body}.
Here, for the intermediate density regime shown in Fig.~\ref{fig:comparison}, we see that different microscopic models of the Bose polaron 
can essentially be collapsed onto the same curve when the ground-state energy is plotted versus $n^{1/3}|a_-|$.
In particular, both the results for the $\Lambda$- and the $r_0$-models are consistent with the QMC calculation from Ref.~\cite{Ardila2016} when Eq.~\eqref{eq:qmc} is used to plot the QMC data in terms of 
the Efimov three-body parameter.
This demonstrates that the polaron energy can be universally described in terms of $n_0^{1/3}|\am|$, regardless of the microscopic details.
Furthermore, it suggests that our variational approach with three Bogoliubov excitations captures the dominant correlations in the full many-body problem, and that the Efimov three-body parameter plays a more important role in the polaron energy than the coherence length of the condensate 
(see Appendix~\ref{app:convergence}).
We emphasize that our comparison in Fig.~\ref{fig:comparison} constitutes a complete reinterpretation of the QMC results of Ref.~\cite{Ardila2016}, 
since that work did not consider the possibility that the variation of the polaron energy with boson-boson interaction was due to Efimov physics.

\section{Characterization of the polaron}

Having shown that the polaron energy is a universal function of
$n^{1/3}|a_-|$ for sufficiently low densities, we now turn to the other quasiparticle properties which characterize the polaron: the residue, effective mass, and contact.
We will generally focus on the $r_0$-model, which is the more physical model and which allows a perturbative analysis in the high-density limit.

\subsection{Residue}

The residue $Z$ 
is the squared overlap of the polaron wave function
with the non-interacting state $c_0^\dag \ket{\Phi}$. 
Making reference to the wave function in Eq.~\eqref{eq:psi},
this is
\begin{align}
Z=|\alpha_0|^2,
\end{align}
where we assume $\left<\Psi|\Psi\right>=1$ for normalization (see 
Appendix~\ref{app:quasi}). The 
residue can be accessed directly in experiment; indeed, in the case of
the Fermi polaron it has been measured using both
radio-frequency spectroscopy~\cite{Schirotzek2009} and Rabi
oscillations~\cite{Kohstall2012}.

\begin{figure}[ht]
\centering
\includegraphics[width=\columnwidth]{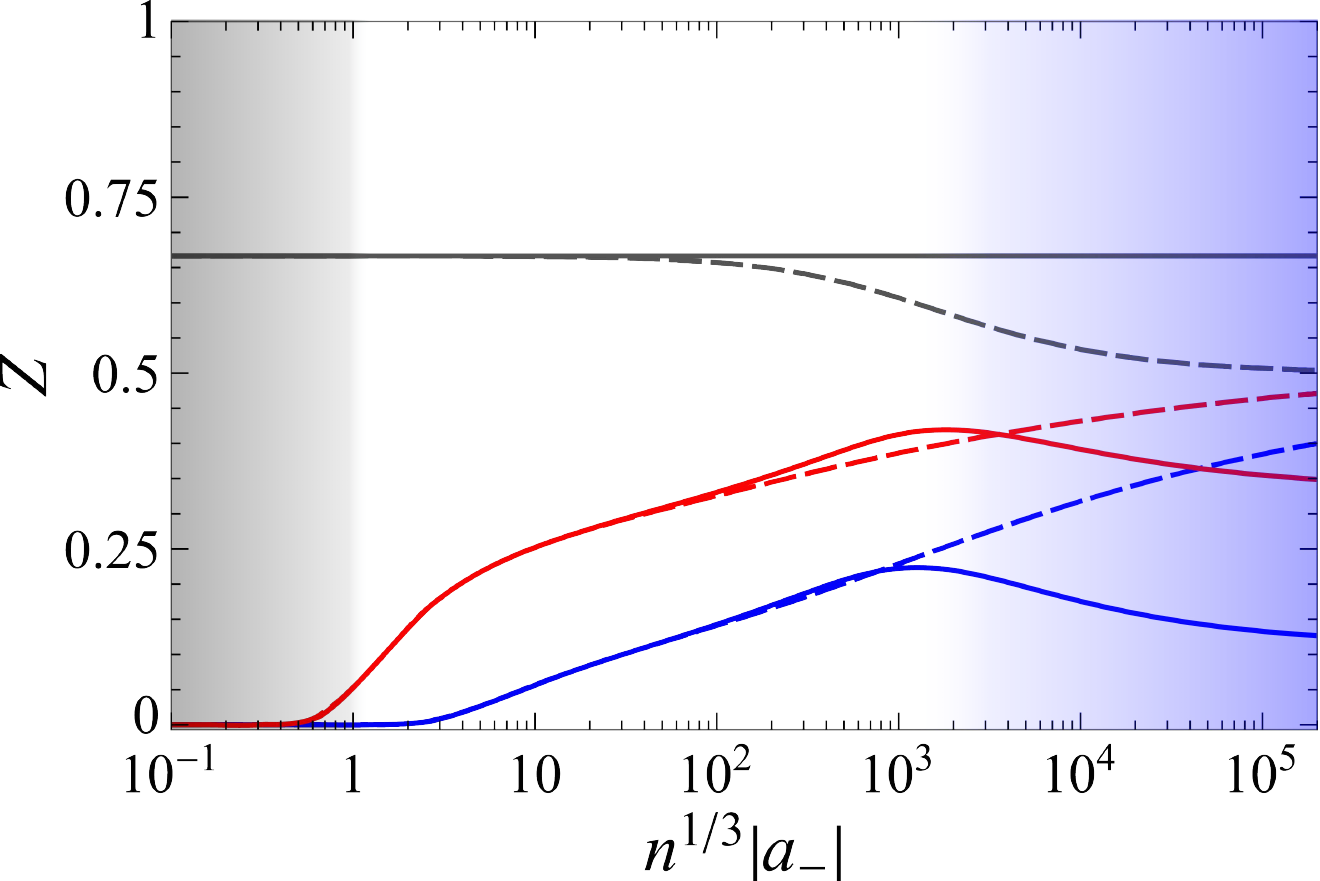}
\caption{Polaron residue $Z$ at unitarity $1/a=0$
as a function of the dimensionless three-body parameter. 
We show the results for the $\Lambda$-model (solid) and the $r_0$-model (dashed) for $\ab \to 0$. In order of decreasing residue, the results are obtained for wave functions including up to one, two or three excitations of the condensate. The grey and blue shaded regions are the same as in Fig.~\ref{fig:universal}, i.e., the low- and high-density regimes, respectively.
}
\label{fig:residue}
\end{figure}

In Fig.~\ref{fig:residue} we show our results for the polaron residue at unitarity
as a function of the three-body parameter, where the boson-boson interactions are taken to be vanishingly small.
We find that there are clear differences
between the results depending on the number of Bogoliubov excitations
used in the variational approach, and also between the different
models. 
Within the $r_0$-model the residue takes the value 1/2 in the high-density regime where the polaron wave function is an equal superposition of the impurity 
and the closed-channel dimer. However, while the residue within the one-excitation approximation is seen to increase from 1/2 to 2/3 with increasing density,
we observe a rapid decrease of $Z$ towards the low-density regime when including several excitations. On the other hand, for the $\Lambda$-model, the residue is 2/3 if we consider only one Bogoliubov excitation. By increasing the number of Bogoliubov excitations, it is suppressed not only in the few-body dominated regime, but also in the high-density limit, where the residue takes the values 1/3 and 1/9, respectively, when two and three excitations are taken into account.
Thus, unlike the energy, the polaron residue does not appear to converge to a universal function of the three-body parameter.

The origin of this behavior
can be understood within
perturbation theory for the $r_0$-model in the high-density limit. Considering again the diagrams in Fig.~\ref{fig:diagrams} and the explicit expression for the self-energy in Eq.~\eqref{eq:sigma}, we find 
at small $g$:
\begin{align} 
Z^{-1} & = 1 - \left.\frac{\partial \Sigma(\0,\omega)}{\partial \omega}\right|_{\omega=E} \nn \\ \label{eq:residue}
& \simeq 2 - \sqrt{\frac{3}{7}} \frac{2}{(8\pi n|r_0|^3)^{1/4}}  + 
\frac{L}{|r_0|},
\end{align}
where $L$ is a low-energy length scale that provides an infrared cutoff of low momenta. In the thermodynamic limit, this is set by the coherence length of the BEC, i.e., we have $L \propto \xi$.
Thus, we find that the residue vanishes when the coherence length $\xi \to \infty$,
and consequently the polaron wave function is orthogonal to the non-interacting impurity state in the case of an ideal BEC. This 
feature cannot properly be captured within 
our variational approach, since this requires an infinite number of Bogoliubov excitations;
hence the observed lack of convergence in Fig.~\ref{fig:residue}.
On the other hand,
when $\ab \neq 0$ we find that the wave functions with two and three excitations both give the same leading order correction to $Z$:
\begin{align}
Z - \frac{1}{2} \propto  \sqrt{\frac{\ab}{|r_0|}} ,
\end{align}
in the limit $n^{1/3}\ab \ll 1$ and $n^{1/3}|r_0| \to \infty$ \footnote{The high-density expansion for $Z$ at finite $\ab$ can, in principle, be calculated using an approach similar to that in Ref.~\cite{Christensen2015}. However, such a lengthy perturbative analysis goes beyond the scope of this work.}.
Thus, in this case, the residue is finite and initially increases with decreasing $|r_0|$.
Note, however, that the residue always vanishes in the low-density regime, as shown in Fig.~\ref{fig:residue}, since the polaron evolves into 
a few-body bound state which has zero overlap with the non-interacting impurity state.
 
The simultaneous convergence of the energy but suppression of the residue in the ideal BEC arises from the fact that the impurity affects the condensate at arbitrarily large distances and thus excites an infinite number of low-energy modes in the condensate. This result is akin to the orthogonality catastrophe for a static impurity in a Fermi gas.
Indeed,  in that case, the orthogonality of the interacting and non-interacting impurity wave functions is also 
apparent in perturbation theory, already at lowest order beyond mean-field theory~\cite{Anderson1995}. We also note that the vanishing residue of the Bose polaron is not confined to the strongly interacting impurity at unitarity: Even for weak impurity-boson interactions, the residue of the attractive polaron was found within perturbation theory to approach zero as $1/\xi$ when $\ab\to0$~\cite{Christensen2015}.

\subsection{Effective mass}
\begin{figure}[ht]
\centering
\includegraphics[width=\columnwidth]{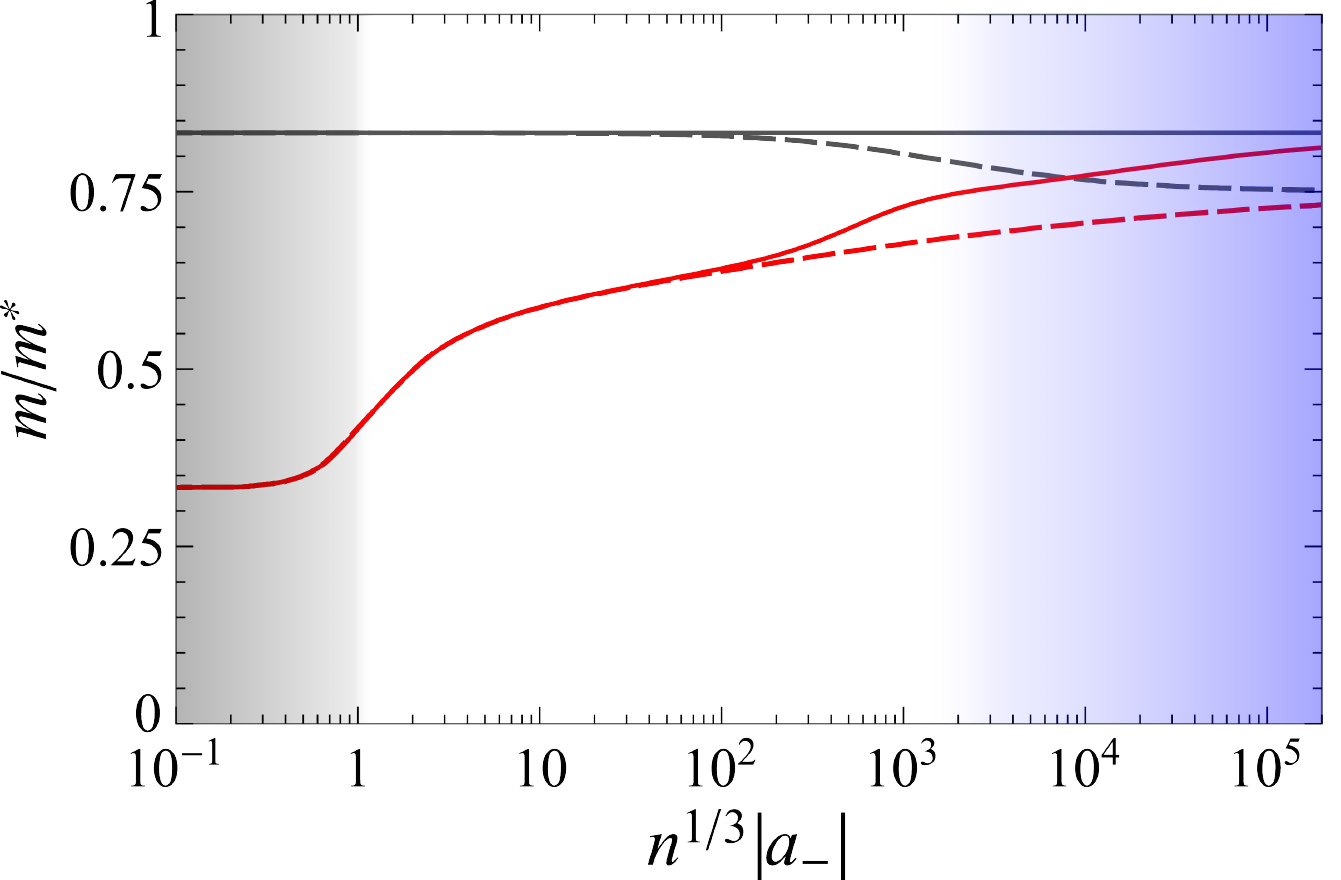}
\caption{Inverse effective mass of the Bose polaron at unitarity, $1/a=0$, for $\ab=0$. We calculate it with the $\Lambda$-model (solid) and the $r_0$-model (dashed), taking into account one (gray) and two (red) Bogoliubov excitations. The grey and blue shaded regions are the same as in Fig.~\ref{fig:universal}
}
\label{fig:effmass}
\end{figure}

The interactions between the impurity and the bosonic medium also modify the mobility of the impurity and give rise to an effective mass $m^*$. 
This can be derived from the impurity's self-energy as follows:
\begin{align} \label{eq:mass}
\frac{m}{m^*} &= Z \left[ 1 + m\left.\frac{\partial^2 \Sigma(\q,E)}{\partial q^2}\right|_{q=0} \right] .
\end{align}
Note that this gives $m^* = m$ for the non-interacting impurity state, as required. Our calculation of $m^*$ within the variational approach is described in Appendix~\ref{app:quasi}, and the results for $\ab \to 0$ at unitarity 
are plotted in Fig.~\ref{fig:effmass}.
Due to the complexity of the calculation, we restrict ourselves to wave functions with up to two Bogoliubov excitations.

For the case of 
an ideal BEC, we have residue $Z \to 0$, and thus one might conclude from Eq.~\eqref{eq:mass} that the effective mass diverges. However, if we once again consider the high-density limit of the $r_0$-model and use the self-energy in Eq.~\eqref{eq:sigma}, we obtain
\begin{align}
  \frac{m}{m^*}  & \simeq Z \left(\frac{3}{2} -  \sqrt{\frac{3}{7}} \frac{1}{(8\pi n|r_0|^3)^{1/4}} + 
  \alpha \frac{L}{|r_0|} 
  \right),
\end{align}
where $\alpha$ is a constant that must be determined from a proper finite-$\ab$ perturbative analysis. 
Thus, we see that there is a term proportional to $L$ that cancels the corresponding term in the residue in Eq.~\eqref{eq:residue}, implying that the effective mass is finite.

Indeed, in the low-density few-body regime, we already have the situation where the residue vanishes, 
while the effective mass remains finite with $m^* = (N+1)m$, where $N$ is the number of bosons bound to to the impurity. This behavior is captured by both $r_0$- and $\Lambda$-models, as shown in Fig.~\ref{fig:effmass}, where the wave function with two Bogoliubov excitations has $m^*/m \to 3$ as $n \to 0$. Likewise, we expect the wave function with three excitations to recover the mass of the tetramer in this limit.
For higher densities, we find that the results of our variational approach become sensitive to the specific model used, as well as to the number of excitations of the ideal BEC considered. Here, in the many-body limit, one requires an infinite number of low-energy excitations to correctly describe the effective mass for $\ab \to 0$, similarly to the case of the residue. 

For a BEC with $0<n^{1/3}\ab \ll 1$, the behavior of the effective mass in the high-density limit of the $r_0$-model can once again be captured using just two Bogoliubov excitations of the BEC. In the limit $|r_0| \to \infty$, the polaron effective mass $m^*/m \to 4/3$, which corresponds to
twice the reduced mass of the impurity and closed-channel dimer.
Moreover, we find that the leading order correction behaves as 
\begin{align}
\frac{m}{m^*} - \frac{3}{4}  \propto \sqrt{\frac{\ab}{|r_0|}} .
\end{align}
Thus, the effective mass in this case initially decreases with decreasing density, before increasing towards the mass of the deepest bound few-body cluster.

\subsection{Contact}
Another quantity of interest is the unitary Tan contact, 
which is defined from the polaron energy as~\cite{Tan_2008a,*Tan_2008b,*Tan_2008c,Braaten2008}
\begin{align} \label{eq:contact}
    C & = m^2 g^2 \left.
    \frac{\partial E}{\partial \nu} \right|_{\nu=0} .
\end{align}
Here we have used the physical detuning  $\nu \equiv - g^2 m/4\pi a$, which can be related to the bare detuning $\nu_0$ using Eq.~\eqref{eq:observables}. 
Physically, the contact gives a measure of the density of pairs at short distances. For the two-channel Hamiltonian in Eq.~\eqref{eq:Ham2}, one can show that the contact is proportional to the population of closed-channel dimers~\cite{Braaten2008}:
\begin{align}
    C 
    &= 
    m^2g^2 \sum_\k \bra{\Psi} d^\dag_\k d_\k \ket{\Psi} .
\end{align}
Within our variational ansatz, this corresponds to
\begin{align}
  C  &= 
    m^2g^2\lba
        |\gamma_0|^2 + \sum_\k |\gamma_\k|^2
        + \frac{1}{2} \sum_{\k_1 \k_2} |\gamma_{\k_1 \k_2}|^2
    \rba .
\end{align}
Note that the limit $g\to\infty$ is well defined, and thus this expression also captures the single-channel $\Lambda$-model.

\begin{figure}[ht]
\centering
\includegraphics[width=\columnwidth]{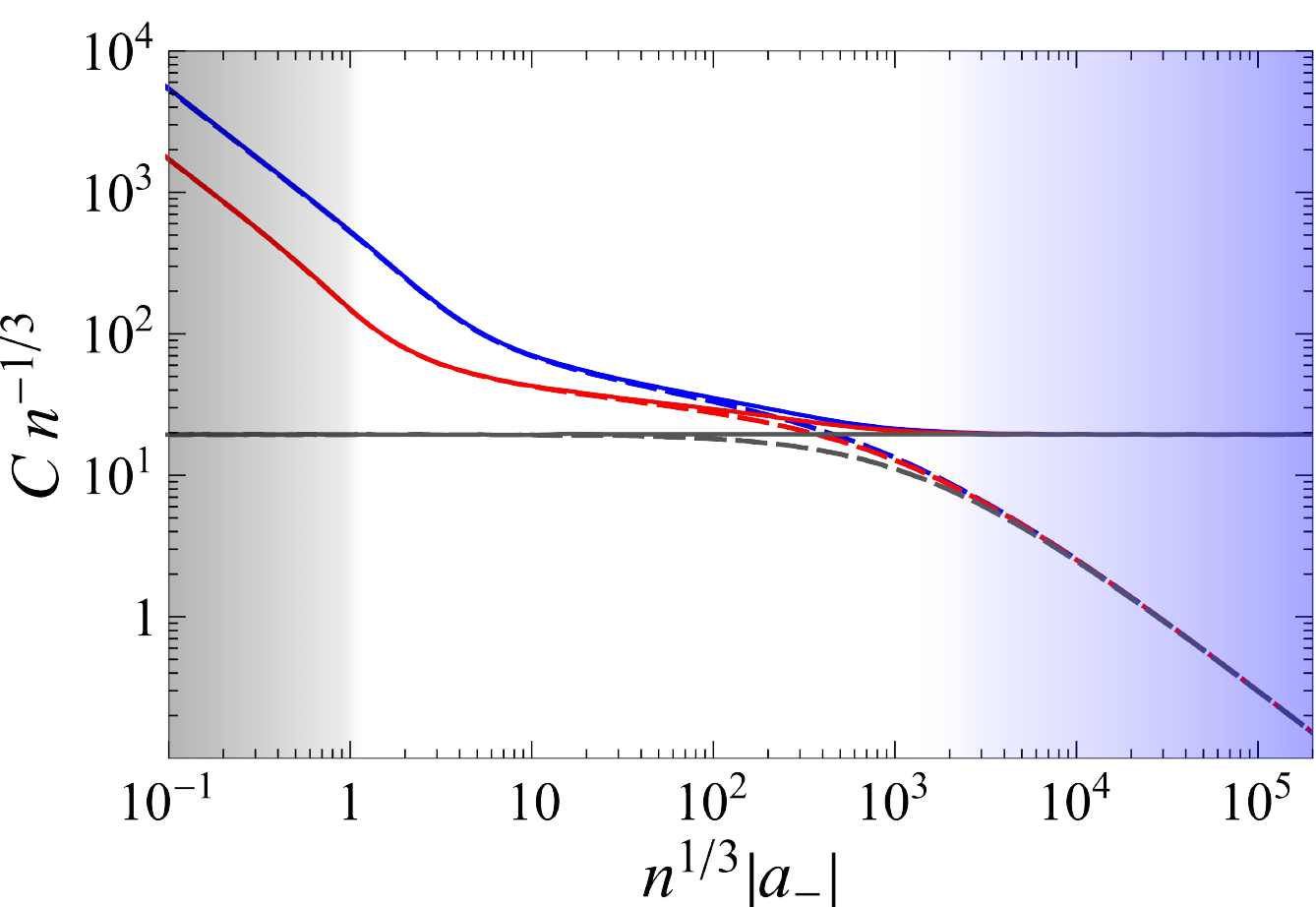}
\caption{Contact of the Bose polaron at unitarity, $1/a=0$, for $\ab=0$. 
The solid and dashed curves correspond to the $\Lambda$- and $r_0$-model, respectively. We show the results obtained using wave functions with up to one (gray), two (red) and three (blue) excitations. The grey and blue shading indicate the same regions as in Fig.~\ref{fig:universal}
}
\label{fig:contact}
\end{figure}

Figure \ref{fig:contact} displays the contact at unitarity for both $\Lambda$- and $r_0$-models in the case of vanishing boson-boson interactions. Similarly to the polaron energy, we find that the contact in the high-density limit is well converged with respect to the number of Bogoliubov excitations.
For the $r_0$-model, we can evaluate the contact perturbatively for large $n^{1/3} |r_0|$ using the self-energy in Eq.~\eqref{eq:sigma}. Here we simply shift the closed-channel dispersion from $\epsilon_\p^d$ to $\epsilon_\p^d + \nu$ and then determine the ground-state energy from the equation $E = \Sigma(\0,E)$ as a function of detuning $\nu$. Applying Eq.~\eqref{eq:contact} then yields the high-density expression
\begin{align}
C  \simeq \frac{8\pi}{|r_0|}\left(\frac{1}2 - \frac{29}{28}\sqrt{\frac{3}{7}}\frac{1}{(8\pi n|r_0|^3)^{1/4}} \right) .
\end{align}
We find that this perturbative result is correctly reproduced by 
both two- and three-excitation approximations in the high-density limit. 
The leading order term is independent of density since the closed-channel fraction tends to $1/2$ at unitarity as $|r_0| \to \infty$ (or, equivalently, as $g\to 0$), see Eq.~\eqref{eq:residue}.

For lower densities $n^{1/3} |a_-| \lesssim 10^3$, the contact becomes insensitive to the underlying model and thus appears to be a universal function of the dimensionless three-body parameter (see Fig.~\ref{fig:contact}). 
In the limit of vanishing density $n \to 0$, we recover the contact of the deepest bound few-body cluster, where $C |a_-|$ is a universal number.

\section{Conclusions and outlook}

Using a variational approach, we have investigated the ground state of
the Bose polaron with unitarity limited impurity-boson
interactions. For sufficiently low boson densities, we have
demonstrated that the polaron equation of state is a universal
function of the three-body parameter associated with Efimov
physics. This is a highly non-trivial result, since the regime of
universality extends to densities at which the binding energies of the
few-body states are several orders of magnitude smaller than the
polaron energy. Our findings are corroborated by the fact that 
we observe the same behavior for two different microscopic models, 
and by
the fact that our calculations agree well
with recent QMC
results~\cite{Ardila2015,Ardila2016} when they are appropriately reinterpreted in terms of 
the three-body parameter.
We have also demonstrated that the few-body spectrum of Efimov trimers and associated tetramers is model independent, further highlighting the universality of our results.

In the non-universal, high-density limit, we have performed a controlled perturbative analysis of the $r_0$-model and we have derived new analytic expressions for the unitary Bose polaron at a narrow Feshbach resonance. Our results demonstrate that even when we take the ideal Bose gas limit, the polaron energy and contact can remain well defined thermodynamically. 
This is a consequence of the fact that only one boson at a time can occupy the closed-channel dimer state, which thus generates an effective repulsion between bosons that acts to stabilize the system. We expect a similar scenario for the $\Lambda$-model since the
high-momentum cut-off in the three- and four-body terms of the
variational equations is equivalent to an
explicit three-body repulsion~\cite{Bedaque1999}. 
However, in general, it is not clear that a well-defined thermodynamic limit exists for treatments of the Bose polaron that focus on single-channel Hamiltonians and neglect Efimov physics~\cite{Shchadilova2014,Grusdt2017}.

While the polaron equation of state at unitarity is relatively insensitive to boson-boson interactions, we find that the quasiparticle residue and effective mass strongly depend on the coherence length of the BEC. In particular, the residue vanishes as $\xi \to \infty$, a feature which resembles the orthogonality catastrophe~\cite{Anderson1967}. 
Thus, an interesting remaining question is how the residue tends to zero with 
the 
number of excitations of the condensate included in the wave function. A related question is: what finite result does the effective mass converge towards as $\xi \to \infty$ and is it a universal function of the three-body parameter?

We emphasize that the (two-channel) $r_0$-model 
faithfully describes the physics of an actual Feshbach resonance in an ultracold atomic gas~\cite{Timmermans1999fri}, and thus we expect our results to accurately describe current and future cold atom experiments.
Furthermore, we expect the universal behavior presented here to also exist for
heavy impurities, a scenario which can be realized by choosing atomic species with a large mass ratio or by pinning the impurity in an optical lattice. 
In the particular case of a static impurity, there are no Efimov few-body states, and therefore it would be of great interest to understand how the system behaves in the limit $n^{1/3}|r_0| \ll 1$ of the two-channel model.

Our results may be probed in experiments similar to those carried out
recently at JILA~\cite{Hu2016} and in Aarhus~\cite{jorgensen2016}. In
particular, our results directly apply to the latter experiment which
used $^{39}$K as both the impurity and host atoms. Furthermore, we estimate that 
$n^{1/3}|a_-|\simeq100$ in that experiment, which indicates that it is
already in a regime where the non-trivial dependendence on the
interparticle spacing is strongly pronounced. Indeed, in a recent JILA experiement~\cite{Klauss2017}, the density of the Bose gas was changed by two orders of magnitude, and thus an experimental investigation of the density dependence of the unitary Bose polaron appears well within reach.

\begin{acknowledgments}
We gratefully acknowledge fruitful discussion with Luis A.~Pe{\~n}a
Ardila, Jan J.~Arlt, Georg M.~Bruun, Rasmus S.~Christensen, Victor Gurarie, Nils B.~J{\o}rgensen, Richard Schmidt, Zhe-Yu Shi, and Bing Zhu. We also thank Luis A.~Pe{\~n}a
Ardila for the QMC data of Ref.~\cite{Ardila2016} and for helpful comments on the manuscript.
SMY acknowledges support from  the Japan Society for the Promotion of Science 
through Program for Leading Graduate Schools (ALPS) and 
Grant-in-Aid for JSPS Fellows (KAKENHI Grant No.~JP16J06706).
JL and MMP acknowledge financial support
from the Australian Research Council via Discovery Project
No.~DP160102739. JL is supported through the Australian
Research Council Future Fellowship FT160100244. 
JL and MMP acknowledge funding from the Universities Australia --
Germany Joint Research Co-operation Scheme.
This work was performed in part at the Aspen Center for Physics, which is supported by the National Science Foundation Grant No.~PHY-1607611.
\end{acknowledgments}

\onecolumngrid
\newpage

\appendix

\section{Coupled integral equations}
\label{app:inteqs}

Taking the derivative $\bra{\partial\Psi} (\hat{H} - E) \ket{\Psi}=0$  with respect to $\alpha_0$, $ \alpha_\k$, $\alpha_{\k_1 \k_2}$, $\alpha_{\k_1 \k_2 \k_3}$, $\gamma_0$, $\gamma_\k$, and $\gamma_{\k_1 \k_2}$ for our variational wave function in Eq.~\eqref{eq:varWF} yields the following linear equations 
\begin{align}
\label{eq:a1}E \alpha_0 = & g \sqrt{n_0} \gamma_0 - g \sum_\k v_\k \gamma_\k, \\
\left(E\!-\! \epsilon_\k \!-\! E_\k  \right) \alpha_\k =  &  g u_\k \gamma_0 +  g\sqrt{n_0} \gamma_\k - g\sum_{\q} \gamma_{\k \q} v_{\q}, \\
\left(E\! - \! E_{\k_1 \k_2}\right) \alpha_{\k_1 \k_2} = &  g \left(\gamma_\vect{k_1}u_\vect{k_2} \! +\! \gamma_\vect{k_2} u_\vect{k_1}  \right) + g\sqrt{n_0}\gamma_{\k_1 \k_2}, \\
\left(E\!- \! E_{\k_1 \k_2 \k_3}\right) \alpha_{\k_1 \k_2 \k_3 } =& g\left(\gamma_{\k_1 \k_2}u_{\k_3}\! + \! \gamma_{\k_2 \k_3}u_{\k_1}\! +\! \gamma_{\k_1 \k_3}u_{\k_2} \right), 
\label{eq:a4}
\\
(E-\nu_0) \gamma_0 = &  g \sqrt{n_0} \alpha_0 + g \sum_\k u_\k \alpha_\k, \\
\left( E- \! \epsilon^{\rm d}_\k \!- \!\nu_0 \!-\! E_\k  \right) \gamma_\k
  = &  g \sqrt{n_0} \alpha_\k - g  v_\k \alpha_0 + g\sum_{\k'} u_{\k'} \alpha_{\k\k'},\\
\left( E- \! \epsilon^{\rm d}_{\k_1 +\k_2} \!- \!\nu_0 \!-\! E_{\k_1}\!-\! E_{\k_1}  \right) \gamma_{\k_1 \k_2}=& g\sqrt{n_0} \alpha_{\k_1 \k_2}-g\left(\alpha_{\k_1 }v_{\k_2}\! + \! \alpha_{\k_2 }u_{\k_1}\!\right) + g\sum_{\k_3} \alpha_{\k_1 \k_2 \k_3 }u_{\k_3 },
\end{align}
where $E_{\k_1 \k_2}=E_{\k_1} \! +\! E_{\k_2} + \epsilon_{\k_1 + \k_2}$ and $E_{\k_1 \k_2 \k_3}=E_{\k_1} \!+\! E_{\k_2}\! + \! E_{\k_3} \!+\! \epsilon_{\k_1 + \k_2 + \k_3}$. Using the first four equations to remove the $\alpha$ coefficients, we obtain
\begin{align}
\T\inv(E,\0) \gamma_0
    =& \frac{n_0}{E} \gamma_0
        + \sqrt{n_0} \sum_{\k} \lba
            \frac{u_\k \gamma_{\k}}{E-\ek-E_\k} - \frac{v_\k \gamma_{\k}}{E}
        \rba  
        -\sum_{\k\q} \frac{u_\k v_\q \gamma_{\k\q}}{E-\ek-E_\k}, \label{eq:2bodyapp} \\
\T\inv(E-E_\k,\k) \gamma_\k
=& \sqrt{n_0} \lba 
        \frac{u_\k}{E-\ek-E_\k} - \frac{v_\k}{E}
    \rba \gamma_0 
   + \frac{n_0}{E-\ek-E_\k} \gamma_\k
    + \sum_{\q} \lba
        \frac{u_\k u_\q  \gamma_\q}{E-E_{\k\q}} 
        + \frac{v_\k v_\q  \gamma_\q}{E}
    \rba \nn \\
&   + \sqrt{n_0} \sum_\q \lba 
        \frac{u_\q \gamma_{\k\q}}{E-E_{\k\q}} - \frac{v_\q  \gamma_{\k\q}}{E-\ek-E_\k} 
    \rba,\label{eq:3bodyapp} \\
\T\inv(E-E_{\k_1}-E_{\k_2},\k_1+\k_2) \gamma_{\k_1 \k_2}
=& \frac{n_0 }{E-E_{\k_1 \k_2}}\gamma_{\k_1 \k_2} \nn \\
& \hspace{-57mm}+ \left[\left\{  \sqrt{n_0} \lba
        \frac{u_{\k_1}\gamma_{\k_2}}{E-E_{\k_1 \k_2}}
        - \frac{v_{\k_1} \gamma_{\k_2}}{E-\ektwo-E_{\k_2}} 
    \rba - \frac{u_{\k_1} v_{\k_2} \gamma_0}{E-\ekone-E_{\k_1}}
  + \sum_{\q} \lba
        \frac{u_{\k_1} u_\q \gamma_{\k_2 \q}}{E-E_{\k_1 \k_2 \q}} 
        + \frac{v_{\k_1} v_\q \gamma_{\k_2 \q}}{E-\ektwo-E_{\k_2}}
    \rba 
   \right\} + (\k_1 \leftrightarrow \k_2)\right]. \label{eq:4bodyapp}
\end{align}
Here, the two-body $T$ matrix $\T(E,\k)$ in the BEC medium is 
\begin{align}
    \T\inv(E,\k)
    &= \frac{m}{4\pi a} - \frac{m r_0}{8\pi} (E-\epsilon^d_{\k})
        - \sum_{\q} \lbc
        \frac{u_\q^2}{E-E_\q-\epsilon_{\k+\q}} + \frac1{2\eq} 
    \rbc \nn \\ & = 
    \T_0\inv(E,\k)
    + \sum_{\q} \lbc
        \frac{1}{E-\eq-\epsilon_{\k+\q}} -
        \frac{u_\q^2}{E-E_\q-\epsilon_{\k+\q}}
    \rbc,
\end{align}
where the vacuum $T$ matrix $\T_0(E,\k)$ was defined in Eq.~\eqref{eq:tmatrix}.

In this paper, we show results obtained from variational wave functions which include 1, 2, or 3 excitations of the condensate, see Eq.~\eqref{eq:psi}. These approximations correspond to solving Eq.~\eqref{eq:2bodyapp} with $\gamma_{\k}=\gamma_{\k\q}=0$, Eqs.~(\ref{eq:2bodyapp}-\ref{eq:3bodyapp}) with $\gamma_{\k\q}=0$, and all of Eqs.~(\ref{eq:2bodyapp}-\ref{eq:4bodyapp}), respectively.

\section{Effect of finite coherence length and number of Bogoliubov excitations}
\label{app:convergence}

\begin{figure}[t]
    \centering
    \includegraphics[width=.5\columnwidth]{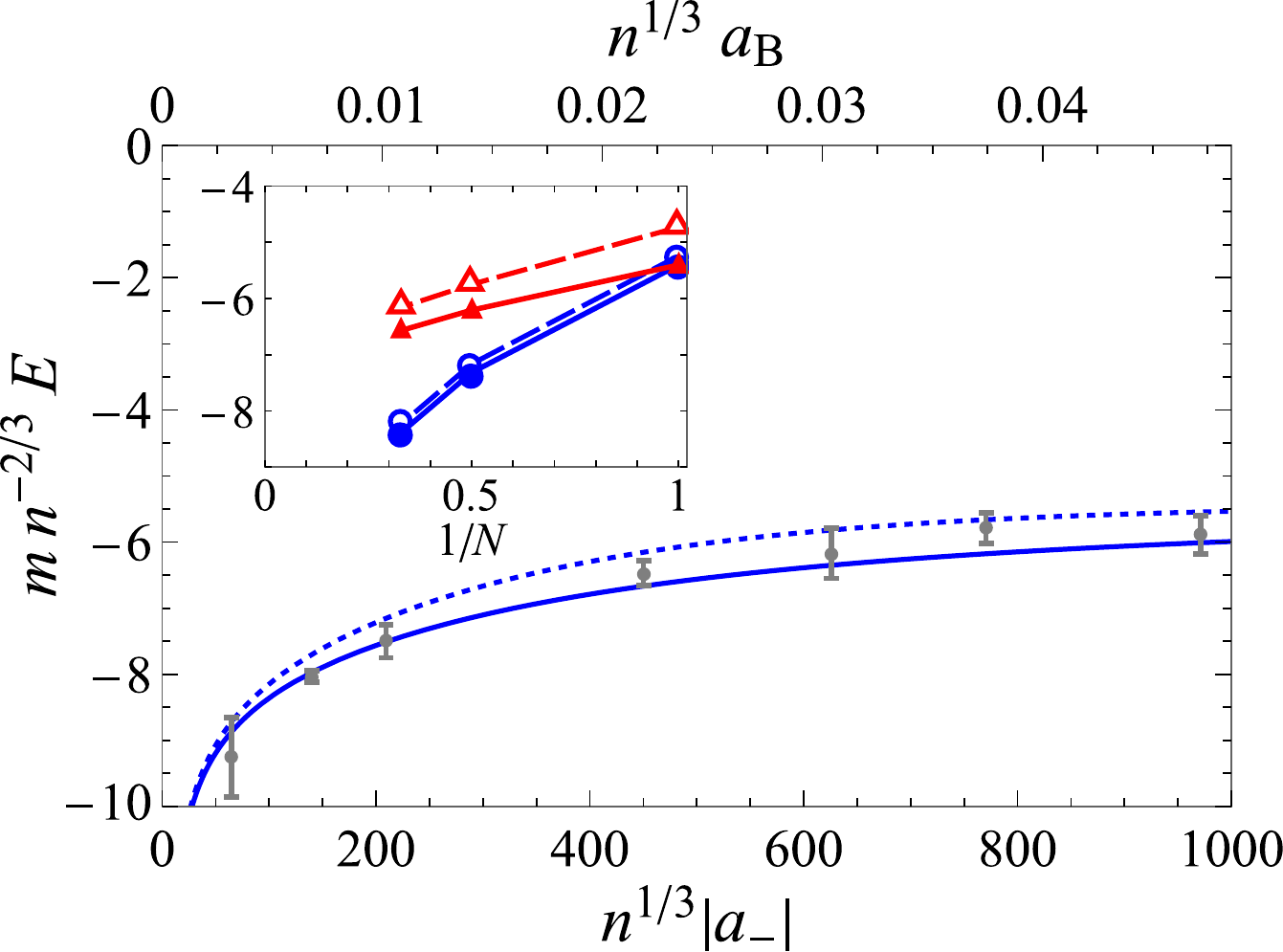}
    \caption{Polaron energy calculated within the $\Lambda$-model as a function of $n^{1/3}|a_-|$ for $\ab=0$ (solid line) and for 
    $\ab=2.1\times10^4|a_-|$. The latter choice of $\ab$ ensures that we take the same ratio of $a_-$ and $\ab$ as in the QMC calculation (gray dots), see Eq.~\eqref{eq:qmc}, and the corresponding scale for $\ab$ is shown on the top $x$-axis. Inset: Polaron energy  at unitarity as a function of inverse number $N$ of Bogoliubov excitations. We present the results of the $\Lambda$-model (solid) and $r_0$-model (open) for $n^{1/3}|a_-|=100$ (blue circles) and $n^{1/3}|a_-|=500$ (red triangles).
resonance, $1/a=0$, for $\ab=0$. 
}
    \label{fig:finite-ab}
\end{figure}

In Section~\ref{sec:manybodyuniversal} we compared the results of our
variational approach with those obtained in quantum Monte Carlo
calculations by Pe{\~n}a Ardila and Giorgini~\cite{Ardila2016}, and
found very good agreement across more than an order of magnitude in
interparticle spacing, see Fig.~\ref{fig:universal}. In this comparison we highlighted the fact that the
dependence of the polaron energy on the three-body parameter $\am$ is much
stronger than the dependence on $\ab$, and to emphasize this we took $\ab=0$. However, as we have argued in
Section~\ref{sec:few-body}, in the QMC calculation the ratio between $\am$ and
$\ab$ was kept fixed at a value of  $a_-^{\mathrm{{\scriptscriptstyle (QMC)}}}\simeq- 2.09(36) \times 10^4 \ab^{\rm{\scriptscriptstyle (QMC)}}$   as identified in
Eq.~\eqref{eq:qmc}. Therefore, we now consider the residual effect of
keeping $\am/\ab$ fixed in the same manner as in the QMC. Our results
for the polaron energy within the $\Lambda$-model are shown in
Fig.~\ref{fig:finite-ab}. We see that the polaron energy is almost
unaffected in the low-density, few-body dominated regime, while $\ab$ becomes
more important with increasing density. However, throughout the range plotted, i.e., the universal many-body regime, the relative
change in energy arising from the change in $n^{1/3}|\am|$ clearly greatly exceeds
that from the increase in $n^{1/3}\ab$. We furthermore see that both sets of
results are consistent with our reinterpretation of the QMC results in
terms of a three-body parameter.

In the inset of Fig.~\ref{fig:finite-ab} we furthermore illustrate the
convergence of the polaron energy  as a
function of (inverse) number of Bogoliubov excitations included in our variational approach. In the many-body dominated regime, at $n^{1/3}|\am|=500$, we see that the convergence appears to be nearly linear, with a slope (and extrapolation to an infinite number of excitations) that is consistent with the difference between the finite $\ab$ dotted line and the QMC results in the main figure. Closer to the few-body dominated regime at $n^{1/3}|\am|=100$ our approach converges more slowly due to the strong few-body correlations induced by the trimer and tetramer bound states. However, it is likely that including four Bogoliubov excitations in our approach to account for the hypothesized pentamer (see Sec.~\ref{sec:lowdensity}) would yield a nearly converged result.

\section{Calculation of observables}
\label{app:quasi}

Once the wave function $\ket{\Psi}$ is obtained,
we can calculate physical quantities of interest.
In the following, we assume that the variational wave function is normalized,
$\langle\Psi|\Psi\rangle=1$. When we solve the renormalized equations, (\ref{eq:2bodyapp}-\ref{eq:4bodyapp}), this implies that
\begin{align}
1=|\alpha_0|^2+\sum_\k|\alpha_\k|^2+\frac12\sum_{\k_1\k_2}|\alpha_{\k_1\k_2}|^2+\frac16\sum_{\k_1\k_2\k_3}|\alpha_{\k_1\k_2\k_3}|^2+
        |\gamma_0|^2 + \sum_\k |\gamma_\k|^2
        + \frac{1}{2} \sum_{\k_1 \k_2} |\gamma_{\k_1 \k_2}|^2.
\end{align}
Since we solve for $\gamma_0$, $\gamma_\k$, and $\gamma_{\k_1\k_2}$, we relate the $\alpha$-coefficients to these through Eqs.~(\ref{eq:a1}-\ref{eq:a4}).

\subsection{Effective mass}

For a system with rotational symmetry, the inverse effective mass 
is defined as
\begin{align}
    \frac{1}{m^*}= \frac{\del^2 E(P)}{\del P^2},
\end{align}
where $E(P)$ is the energy dispersion of the Bose polaron. 
The variational ansatz can be easily extended to a Bose polaron 
with a finite momentum, but 
solving the resulting integral equations becomes numerically 
expensive because the equations do not have rotational symmetry. 
Here, following a similar line of argument to Ref.~\cite{Trefzger2012},
we present a method to calculate the effective mass without
solving the integral equations for a finite-momentum Bose polaron.

In general, the coupled integral equations for $\gamma$'s for a finite momentum polaron
have the following structure (similar to Eqs.~(\ref{eq:a1}-\ref{eq:a4})):
\begin{align}
    \hat{X}[E(P),\P] \ket{\gamma} = 0.
\end{align}
Here, $\ket{\gamma}$ is an ordered set of $\gamma_0$, $\gamma_\k$ and 
$\gamma_{\k_1 \k_2}$, and the center-of-mass momentum can be taken to be $P{\bf e}_z$ 
without loss of generality.
While the matrix $\hat{X}$ is not the Hamiltonian, it is still Hermitian.
The energy dispersion $E(P)$ is determined by the zero-crossing of the lowest
(or highest, depending on the sign convention) eigenvalue of $\hat{X}$.
Therefore, for a small $P$, we can expand $\hat{X}$ up to ${\cal O}(P^2)$
and evaluate the lowest eigenvalue in the same way as the Rayleigh-\sch
perturbation theory.
It is straightforward to show that
\begin{align}
    \hat{X}[E(P),P] = \hat{X}_0 + P \hat{X}_P + \frac{P^2}{2} \left(
        \hat{X}_{PP} + 
        (1/m^*)\hat{X}_{E}
    \right) + {\cal O}(P^3),
\end{align}
where
\begin{gather}
    \hat{X}_0 = \hat{X}[E(0),0], \quad
    \hat{X}_P = \left.\frac{\del\hat{X}}{\del P} [E,P]\right|_{_{E=E(0)}^{P=0}}, \quad
    \hat{X}_{PP} = \left.\frac{\del^2\hat{X}}{\del P^2} [E,P]\right|_{_{E=E(0)}^{P=0}}, \quad
    \hat{X}_E = \left.\frac{\del\hat{X}}{\del E} [E,P]\right|_{_{E=E(0)}^{P=0}}.
\end{gather}
Then the lowest eigenvalue $\lambda(P)$ of $\hat{X}[E(P),P]$ can be expressed
as a series expansion in terms of $P$,
\begin{align}
    \lambda(P) = \lambda_0 + \lambda_1 P + \lambda_2 P^2 + {\cal O}(P^3),
\end{align}
and its coefficients $\lambda_i$ are determined perturbatively:
\begin{align}
    \lambda_1 &= \bra{\gamma^{(0)}} \hat{X}_P \ket{\gamma^{(0)}} = 0, \\
    \lambda_2 &= \frac{1}{2} \bra{\gamma^{(0)}} \left(
        \hat{X}_{PP} + (1/m^*)
        \hat{X}_E
    \right) \ket{\gamma^{(0)}}
    + \sum_{i>0} 
        \frac{|\bra{\gamma^{(i)}}\hat{X}_P\ket{\gamma^{(0)}}|^2}
        {\lambda_0^{(0)}-\lambda_0^{(i)}}.
\end{align}
Here, $\lambda^{(i)}_0$ and $\ket{\gamma^{(i)}}$ are the $i$-th eigenvalue and 
its corresponding eigenvector of $\hat{X}_0$, and $i=0$ corresponds to the state that satisfies $\hat X_0[E(0),\0]\ket{\gamma^{(0)}}=\lambda_0^{(0)}\ket{\gamma^{(0)}}=0$.
$\lambda_1=0$ follows from the rotational invariance of $\ket{\gamma^{(0)}}$.
Now, recalling that the energy dispersion is found by zero-crossing of the lowest
eigenvalue of $\hat{X}[E(P),P]$, we can set
\begin{align}
    \lambda(P) = \lambda_0^{(0)} = 0,
\end{align}
which in turn, implies that $\lambda_2 = 0$.
From this, the inverse effective mass 
is obtained as follows:
\begin{align}
1/m^*    &= \bra{\gamma^{(0)}}\hat{X}_E\ket{\gamma^{(0)}}\inv \left[
        \sum_{i>0} \frac{2}{\lambda^{(i)}_0}
            \left| \bra{\gamma^{(i)}}\hat{X}_P\ket{\gamma^{(0)}} \right|^2
        - \bra{\gamma^{(0)}}\hat{X}_{PP}\ket{\gamma^{(0)}}
    \right] \\
    &= \bra{\gamma^{(0)}}\hat{X}_E\ket{\gamma^{(0)}}\inv \left[
            2 \bra{\gamma^{(0)}} \hat{X}_P \hat{Q} \hat{X}_0\inv 
                \hat{Q} \hat{X}_P \ket{\gamma^{(0)}} 
        - \bra{\gamma^{(0)}}\hat{X}_{PP}\ket{\gamma^{(0)}}
    \right],
\end{align}
where $\hat{Q}=1-\sum_{i>0} \ket{\gamma^{(i)}}\bra{\gamma^{(i)}}$.

\twocolumngrid

\bibliography{bosepolaron}

\end{document}